\documentclass[12pt,preprint]{aastex}

\usepackage{graphicx}
\usepackage{rotating}

\shorttitle{}
\shortauthors{Nesvorn\'y et al.}

\begin{document}
\baselineskip 19.pt

\title{NEOMOD 3: The Debiased Size Distribution of Near Earth Objects}

\author{David Nesvorn\'y$^1$, David Vokrouhlick\'y$^2$, Frank Shelly$^3$, Rogerio Deienno$^1$,\\ 
William F. Bottke$^1$, Carson Fuls$^3$, Robert Jedicke$^4$, Shantanu~Naidu$^5$, 
Steven~R.~Chesley$^5$, Paul W. Chodas$^5$, Davide Farnocchia$^5$, Marco Delbo$^6$} 
\affil{(1) Department of Space Studies, Southwest Research Institute, 1050 Walnut St., 
  Suite 300,  Boulder, CO 80302, USA}
\affil{(2) Institute of Astronomy, Charles University, V Hole\v{s}ovi\v{c}k\'ach 2, CZ–18000 
Prague 8, Czech Republic}
\affil{(3) Lunar and Planetary Laboratory, The University of Arizona, 1629 E. University Blvd. 
Tucson, AZ 85721-0092, USA}
\affil{(4) Institute for Astronomy, University of Hawaii, 2680 Woodlawn Drive, Honolulu, HI
  96822-1839, USA}
\affil{(5) Jet Propulsion Laboratory, California Institute of Technology, 4800 Oak Grove Dr.
Pasadena, CA 91109, USA}
\affil{(6) Laboratoire Lagrange, UMR7293, Universit\'e C\^ote d'Azur, CNRS, Observatoire de la 
C\^ote d'Azur, Bouldervard de l'Observatoire, 06304, Nice Cedex 4, France}

\begin{abstract}
Our previous model (NEOMOD2) for the orbital and absolute magnitude distribution of Near Earth Objects 
(NEOs) was calibrated on the Catalina Sky Survey observations between 2013 and 2022. Here we 
extend NEOMOD2 to include visible albedo information from the Wide-Field Infrared Survey Explorer. 
The debiased albedo distribution of NEOs can be approximated by the sum of two Rayleigh distributions 
with the scale parameters $p_{\rm V,dark}\simeq 0.03$ and $p_{\rm V,bright} \simeq 0.17$. We find evidence
for smaller NEOs having (on average) higher albedos than larger NEOs; this is likely a consequence 
of the size-dependent sampling of different main belt sources. These inferences and the absolute magnitude 
distribution from NEOMOD2 are used to construct the debiased size distribution of NEOs. We estimate
$830\pm60$ NEOs with diameters $D>1$ km and $20,\!000 \pm 2,\!000$ NEOs with $D>140$ m. The new model, 
NEOMOD3, is available via the NEOMOD Simulator -- an easy-to-operate code that can be used to 
generate user-defined samples (orbits, sizes and albedos) from the model.   
\end{abstract}

\section{Introduction}

An accurate knowledge of the size distribution of NEOs is interesting for many different reasons, 
including the objectives of the NASA's Planetary Defense Coordination Office 
(PDCO).\footnote{\texttt{https://science.nasa.gov/planetary-defense}} Several 
size-distribution models of NEOs have been developed (e.g., Mainzer et al. 2011, Morbidelli 
et al. 2020, Harris \& Chodas 2021). Mainzer et al. (2011) combined the albedo measurements 
from the cryogenic portion of the Wide-Field Infrared Survey Explorer (WISE)
mission with the magnitude distribution of known NEOs, approximately 
accounting for the incompleteness, to estimate $981\pm19$ NEOs with $D>1$ km and $20,\!500\pm3,\!000$ 
NEOs with $D>100$ m. The albedo distribution was inferred from NEOs detected by WISE, which 
was an appropriate choice because the WISE sample is much less biased with respect to visible albedo 
than surveys in visible wavelengths.

Morbidelli et al. (2020) developed an approximate debiasing method, combined the cryogenic WISE albedos 
with the NEO model from Granvik et al. (2018), and inferred $\sim 1000$ NEOs with $D>1$ km. The strength
of this work relative to Mainzer et al. (2011) was that it used the {\it debiased} orbital and 
absolute-magnitude distribution model (Granvik et al. 2018; also see Bottke et al. 2002) -- this removed uncertainties related to 
the completeness of the known NEO population considered in Mainzer et al. (2011). Morbidelli et al. (2020), 
however, used a relatively crude albedo binning (three bins with $p_{\rm V}<0.1$, $0.1<p_{\rm V}<0.3$ and 
$p_{\rm V}>0.3$; a uniform distribution assumed in each bin), which did not allow them to reconstruct 
the debiased albedo distribution in detail. The inferences given in that work for the size distribution 
of NEOs were therefore somewhat uncertain. 

Finally, Harris \& Chodas (2021) updated their previous model for the absolute magnitude distribution 
of NEOs (Harris \& D'Abramo 2015). A reference albedo $p_{\rm V,ref}=0.14$ (Stuart \& Binzel 2004) was 
used to convert the absolute magnitude distribution into the size distribution. This is less than ideal 
because NEOs have a wide range of visible albedos and it is therefore not obvious if there is a single albedo 
value that can be used to convert the distributions, and if so, what reference albedo should be used (Morbidelli 
et al. 2020 proposed $p_{\rm V,ref}=0.147$). 

Here we combine the absolute magnitude distribution from NEOMOD2 (Nesvorn\'y et al. 2024; 
hereafter Paper II) with the visible albedo information from WISE (Mainzer et al. 2011) to obtain the 
size distribution of NEOs.          

NEOMOD is an orbital and absolute magnitude model of NEOs (Nesvorn\'y et al. 2023; hereafter Paper I).
To develop NEOMOD, we closely followed the methodology from previous studies (Bottke et al. 2002, 
Granvik et al. 2018), and improved it when possible. Massive numerical integrations were 
performed for asteroid orbits escaping from eleven main belt sources. Comets were included as the twelfth 
source. The integrations were used to compute the probability density functions (PDFs) that define 
the orbital distribution of NEOs (perihelion distance $q<1.3$ au, $a<4.2$ au) from each source. We
developed a new method to accurately calculate biases of NEO surveys and applied it to the Catalina 
Sky Survey (CSS; Christensen et al. 2012) in an extended magnitude range ($15<H<28$). The 
\texttt{MultiNest} code (Feroz \& Hobson 2008, Feroz et al. 2009) was used to optimize the (biased) model 
fit to CSS detections. The improvements included: (i) cubic splines to represent the magnitude distribution 
of NEOs, (ii) a physical model for disruption 
of NEOs at low perihelion distances (Granvik et al. 2016), (iii) an accurate estimate of the impact 
fluxes on the terrestrial planets, and (iv) a flexible setup that can be readily adapted to any 
current or future NEO survey. In Paper II (Nesvorn\'y et al. 2024) we extended NEOMOD to incorporate 
new data from CSS.\footnote{The camera of G96 (Mount Lemmon Observatory) was upgraded to a wider field of view 
(FoV; $2.23^\circ \times 2.23^\circ$) in May 2016 and the G96 telescope detected 11,934 unique NEOs between 
May 31, 2016 and June 29, 2022. This can be compared to only 2,987 unique NEO detections of G96 for 
2005--2012 ($1.1^\circ \times 1.1^\circ$ FoVs).} 

Here we upgrade NEOMOD2 to include the WISE data. The main goal is to obtain an accurate estimate of the \textit{size 
distribution} of NEOs. A straightforward approach to this problem would be to use the WISE measurements of NEO 
diameters, develop a debiasing procedure, and infer the size distribution from the WISE data alone. 
During the cryogenic portion of the mission, however, WISE only detected 428 unique NEOs (Mainzer et al. 
2011), which can be compared to over $\sim 15,\!000$ unique NEO detections by CSS between 2013 and 2022. The 
results of the direct approach to this problem, as described above, would therefore suffer from (relatively) 
small number statistics. For this reason, it is better to use the WISE measurements of visible albedo of NEOs, 
debias them, and combine the results with the absolute magnitude distribution from NEOMOD2. This hybrid 
method takes advantage of the full statistics from CSS and the realistic albedo distribution from 
WISE.\footnote{We considered using the {\it Spitzer} observations of NEOs (Trilling et al. 2020) but found 
it difficult to accurately model the observational biases involved in those observations. This is because
NEOs observed by {\it Spitzer} were selected based on their visual magnitudes. The {\it Spitzer} sample 
of NEOs is therefore biased toward high albedos, especially for small NEOs. It was not clear to us how to 
remove this bias because the selected NEOs were discovered by different NEO surveys with different biases.}
  
We test several models with different parameters. The simple model and its variant with the size-dependent albedo
distribution, as described in section 4.1, have fewer parameters and are therefore presumably more robust. We use 
these models to obtain population estimates and impact fluxes. The simple model cannot account for potential 
dependences of the albedo distribution on NEO orbit (e.g., outer main-belt sources may be producing more dark 
NEOs than the inner main-belt sources). We therefore develop a complex model where different NEO sources have 
different contribution to NEOs with low and high albedos (Sect. 4.2; Morbidelli et al. 2020). The complex model 
correctly reproduces the correlation of albedo with orbit inferred from the NEOWISE data, but it has more parameters, 
and at least in some cases \texttt{MultiNest} struggles to constrain them (e.g., the case of Phocaeas; Sect 4.2). 
The NEOWISE statistics with only 428 detections during the cryogenic part of the mission (Mainzer et al. 2011) may 
be not large enough for the complex model to fully converge to a perfect solution. In this situation, we find 
it best to stay conservative and report a relatively large range of estimates that contains the results of all 
explored models. Estimates for NEOs with diameters $D<100$ m are subject to additional uncertainties, as 
the albedo distribution for $D<100$ m needs to be extrapolated from the NEOWISE data for $D>100$ m. 

\section{The base model from Paper II}

In NEOMOD2, the biased NEO model is defined as 
\begin{equation}
{\cal M}_{\rm b}(a,e,i,H) =  n(H)\, {\cal P}(a,e,i,H) \sum_{j=1}^{n_{\rm s}} \alpha_j(H)\, p_{q^*,j}(a,e,i,H)\, \ , 
\label{model}
\end{equation}
where $n(H)$ is the differential absolute-magnitude distribution of the NEO population, 
${\cal P}(a,e,i,H)$ is the CSS's detection probability, $\alpha_j$ are the magnitude-dependent 
weights of different sources ($\sum_j \alpha_j(H) = 1$), $n_{\rm s}$ is the number of NEO sources, 
$p_{q^*,j}(a,e,i,H)$ is the PDF of the orbital distribution of NEOs from the source $j$, 
including the size-dependent disruption at the perihelion distance $q$ (Paper I).

The model domain in $a,e,i,H$ is divided into bins (see Table 2 in Paper I). To determine the survey's 
detection probability in each bin, we place a large number of test bodies in each bin, assume random 
orbital longitudes, and test whether individual bodies are or are not detected. This includes considerations 
related to the geometric bias (i.e., will an object appear in survey's fields of view?), photometric 
sensitivity and trailing loss (Paper II). ${\cal P}(a,e,i,H)$ is then calculated as the mean probability 
that an object with $a,e,i,H$ will be detected over the whole duration of the survey. The orbital 
distributions $p_{q^*,j}(a,e,i,H)$ are obtained from numerical integrations described in Paper I. 
The distributions are normalized such that $\int p_{q^*,j}(a,e,i,H)\ {\rm d}a\, {\rm d}e\, {\rm d}i = 1$ for 
any $H$.

There are three sets of model parameters in NEOMOD2: the (1) coefficients $\alpha_j$, (2) parameters related 
to the absolute magnitude distribution of NEOs, and (3) priors that define the disruption model (Granvik 
et al. 2016). As for (1), we have $n_s=12$ sources in total: eight individual resonances ($\nu_6$, 3:1, 
5:2, 7:3, 8:3, 9:4, 11:5 and 2:1), weak resonances in the inner belt, two high-inclination sources 
(Hungarias and Phocaeas), and comets.\footnote{Note that all comets, including the short- and long-period
comets, were included in NEOMOD and NEOMOD2. The Jupiter-family comets represent the dominant part of 
cometary NEOs with short orbital periods (here $a<3.5$ au).} 
The intrinsic orbital distribution of model NEOs is obtained 
by combining all sources. The coefficients $\alpha_j$ represent the relative contribution of each source 
to the NEO population ($\sum_{j=1}^{n_{\rm s}} \alpha_j = 1$). As the contribution of different sources to NEOs
is size dependent (Papers I and II), $\alpha_j$ are functions of absolute magnitude; we adopt a linear
dependence for simplicity. As for (2), the differential and cumulative absolute magnitude distributions 
are denoted $n(H)={\rm d}N/{\rm d}H$ and $N(H)$, respectively. 
We use cubic splines to represent $\log_{10} N(H)$ (Paper I). As for (3),
we eliminate test bodies when they reach the critical distance $q^*(H)$. We assume 
that the $q^*$ dependence on $H$ is (roughly) linear, and parameterize it by 
$q^* = q_0^* + \delta q^* (H-H_q)$, where $H_q=20$ (the choice of $H_q$ is arbitrary; $q_0^*$ and 
$\delta q^*$ are the model parameters).

The {\tt MultiNest} code is used to perform the model selection, parameter estimation and error analysis 
(Feroz \& Hobson 2008, Feroz et al. 2009).\footnote{\url{https://github.com/farhanferoz/MultiNest}} For 
each \texttt{MultiNest} trial, Eq.~(\ref{model}) is constructed by the methods described above. 
This defines the expected number of events $\lambda_j = {\cal M}_{\rm b}(a,e,i,H)$ in every bin of the model 
domain, and allows \texttt{MultiNest} to evaluate the log-likelihood
\begin{equation}
{\cal L} = - \sum_j \lambda_j + \sum_j n_j \ln \lambda_j \ ,
\label{like}
\end{equation}
where $n_j$ is the number of objects detected by CSS in the bin $j$, $\lambda_j$ is the number of objects 
in the bin $j$ expected from the biased model, and the sum is executed over all bins in $a$, $e$, 
$i$ and $H$ (Paper I). There are 30 model parameters in total: 22 coefficients 
$\alpha_j$,\footnote{To define the linear dependence of $\alpha_j$ on $H$, we define two sets 
of $\alpha_j$ coefficients for bright and faint NEOs, and linearly interpolate between them.
For $n_{\rm s}=12$ sources, this represents 11 coefficients at the bright end (the contribution
of the last source can be computed from $\sum_j \alpha_j(H) = 1$) and 11 coefficients on the faint 
end.} 6 parameters that define the magnitude distribution from splines (five slopes and the overall 
normalization), and 2 parameters for the size-dependent disruption ($q_0^*$ and $\delta q^*$).

\begin{figure}
\epsscale{0.6}
\plotone{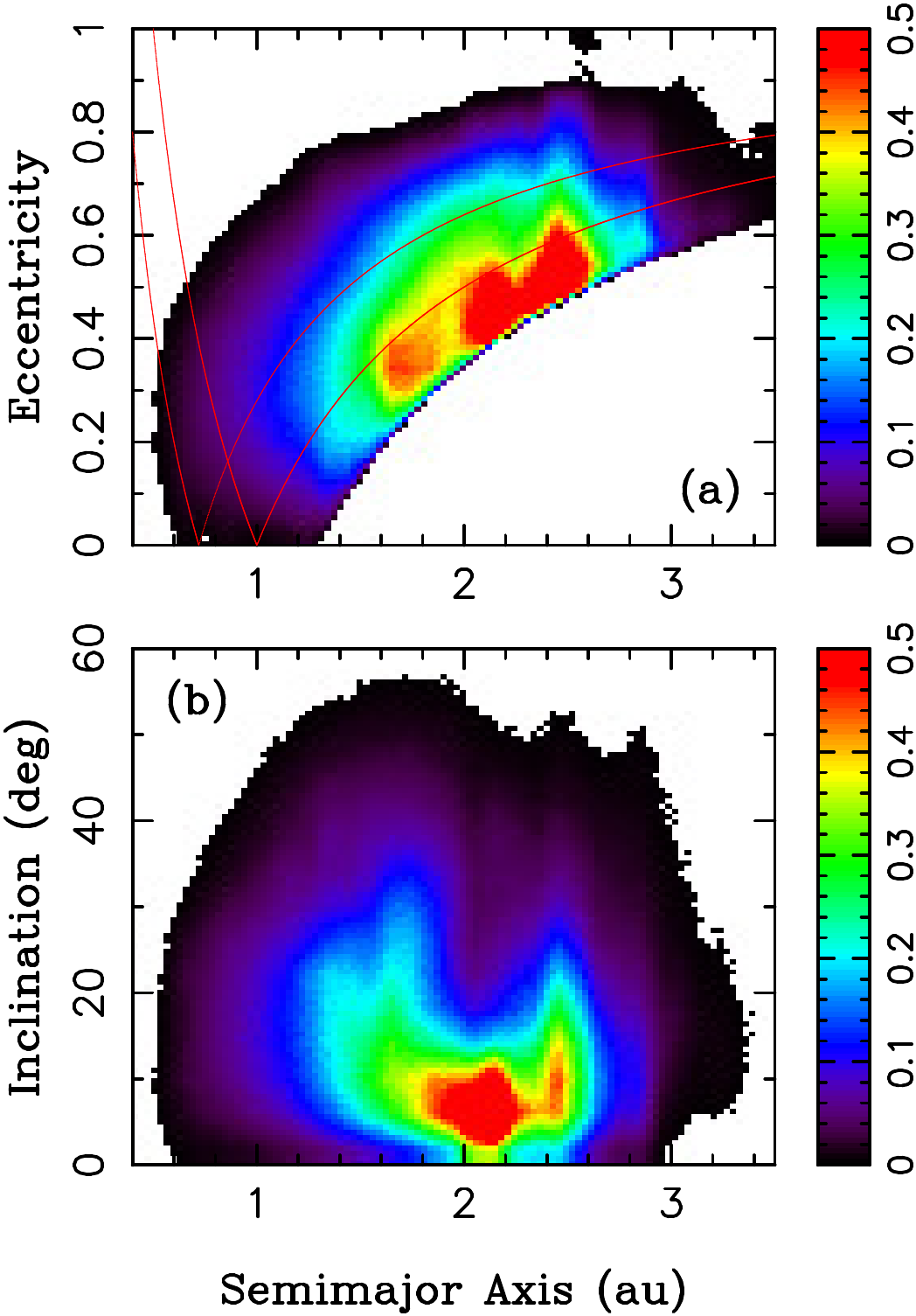}
\caption{The orbital distribution of NEOs from our {\it intrinsic} (debiased) model (NEOMOD2). We used 
the NEOMOD Simulator (Paper II) and generated $1.1\times10^6$ NEOs with $15<H<28$. 
The distribution was marginalized over absolute magnitude and binned using 100 bins in each orbital element 
($0.4<a<3.5$ au, $e<1$ and $i<60^\circ$). Warmer colors correspond to orbits where NEOs are more likely
to reside. In the plot shown here, the maximum residence probability in a bin is normalized to 1.}
\label{unb}
\end{figure}

Once {\tt MultiNest} converges, the maximum likelihood parameters can be used to define the intrinsic 
(debiased) NEO model
\begin{equation}
{\cal M}(a,e,i,H) =  n(H)\, \sum_{j=1}^{n_s} \alpha_j(H)\, p_{q^*,j}(a,e,i,H)\, \ . 
\label{model2}
\end{equation}
Figures \ref{unb} and \ref{harris} show the orbital and absolute magnitude distributions from NEOMOD2.
The orbital distribution in Fig. \ref{unb} is consistent with the NEO model from Granvik et al. (2018).
The absolute magnitude distribution in Fig. \ref{harris} is similar to the one reported in 
Harris \& Chodas (2021, 2023) for $H<25$, but shows a shallower slope and fewer NEOs for $H>25$ (see 
Paper II for a discussion). It has to be noted that the distribution presented in Harris \& Chodas 
(2021, 2023) assumed fixed slopes for $H > 26$. This is because there is a statistially insignificant 
number of re-detection for $H > 26$ and the re-detection method does not give useful results for
these faint magnitudes.

\section{Methods}

\subsection{NEO detections by cryogenic NEOWISE}
WISE is a NASA mission designed to survey the entire sky in four infrared wavelengths: 3.4, 4.6, 12 
and 22 $\mu$m, denoted $W1$, $W2$, $W3$ and $W4$, respectively (Mainzer et al. 2005, Liu et al. 2008).
The survey began on January 14, 2010. The mission exhausted its primary tank cryogen on August 5, 
2010 and secondary tank cryogen on October 1, 2010. An augmentation to the WISE processing pipeline,
NEOWISE, permitted a search and characterization of moving objects. The survey has yielded observations 
of over 157,000 minor planets, including NEOs, main belt asteroids, comets, Trojans, Centaurs and 
Kuiper belt objects (Mainzer et al. 2011). The survey was continued as the NEOWISE Post-Cryogenic Mission 
using only bands $W1$ and $W2$. 

For the purposes of determining the debiased population of NEOs, in this paper, we only consider NEOs  
detected during the fully cryogenic portion of WISE. This data set consists of 428 
NEOs (Fig. \ref{baloons}), of which 314 
were rediscoveries of objects known previously and 114 were NEOWISE discoveries (Mainzer et al. 2011). 
The ranges of visual albedos, diameters and absolute magnitudes of NEOs detected by 
NEOWISE are $0.01 <  p_{\rm V} \lesssim  0.5$, $0.1 < D < 10$ km and $13 \lesssim H \lesssim 23$, respectively.

The non-cryogenic portion of WISE is not considered here, because the $W1$ and 
$W2$ bands mix the reflected light with thermal emission, and are less useful for accurate albedo 
determinations. The cryogenic NEOWISE sample is only weakly biased with respect to visible albedo. For 
comparison, a survey in visible wavelengths such as CSS typically detects objects to some limiting 
apparent magnitude $V_{\rm lim}$. This results in a magnitude-limited sample where the population is 
characterized to some faint absolute magnitude limit, $H_{\rm lim}$; bodies with low visual albedos 
can be severely underrepresented for $H<H_{\rm lim}$ (Appendix A).

\subsection{Thermal infrared bias}
The intrinsic albedo distribution of NEOs is close but not exactly equal to that of NEOs detected 
by NEOWISE. This is because objects with low visible albedo absorb more sunlight and emit 
more thermal radiation; they are therefore more easily detected in infrared wavelengths. The NEOWISE 
sample is thus (slightly) biased toward NEOs with low visual albedos. This is only a modest effect
for $D>1$ km (Mainzer et al. 2011), because large bodies with low and high albedos were detected 
nearly equally well by NEOWISE, but it can become increasingly important for $D<1$ km NEOs for which
the thermal emission in the $W3$ band can be weak.
 
We used the Near-Earth Asteroid Thermal Model (NEATM) model (Harris 1998) to account for the thermal 
infrared bias. NEATM adopts several simplifying assumptions. Objects are assumed to be perfectly spherical. 
NEATM does not physically account for thermal inertia -- it empirically models it using the beaming 
parameter, $\eta$. Mainzer et al. (2011) fitted $\eta$ for 313 NEOs with measurements in two or more 
thermal bands and found the median value $\eta=1.4$. We tested different values of $\eta$ in a 
0.4 range around $\eta=1.4$ and found that the results are not sensitive to this choice. We therefore 
adopted $\eta=1.4$ as a fiducial value. The color corrections from Wright et al. (2010) were applied. 

Here we model NEOWISE detections in the $W3$ band, which was available only during the cryogenic portion of 
the WISE mission, and had better sensitivity than the $W4$ band (surface temperatures of NEOs imply
peak black body emission near the center of $W3$). The detection in the $W3$ band is therefore 
a good proxy for NEO detection by NEOWISE {\it and} a reliable measurement of asteroid albedo. The photometric 
detection probability of NEOWISE as a function of $W3$ magnitude was obtained as a ratio of detected 
and available NEOs in Mainzer et al. (2011). Adopting their Eq. (3), we have 
\begin{equation}
P(W3)={\epsilon_0 \over 1 + \exp \left ( { W3 - W3_{\rm lim} \over W3_{\rm wid} } \right ) } 
\label{w3}
\end{equation}   
with $\epsilon_0 = 0.9$, $W3_{\rm lim}=10.25$ and $W3_{\rm wid}=0.2$. This is the same functional form
that we used to model CSS detections in the apparent magnitude $V$ in Paper I. The parameters
$\epsilon_0$,  $W3_{\rm lim}$ and $W3_{\rm wid}$ were fixed to provide the best fit to the median detection 
probabilities shown in Fig. 11 in Mainzer et al. (2011). We verified that small changes of these 
parameters do not substantially affect the results reported here. 

To understand the thermal infrared bias in detail, we used the NEOMOD
simulator (Paper II) and generated orbital elements $a$, $e$ and $i$ of $10^5$ model NEOs. The 
orbits were given a uniformly random distribution of orbital longitudes. For each diameter set, 
all bodies were assigned the same value of visible albedo $p_{\rm V}$ and the detection probabilities 
in the $W3$ band were computed individually for them. To respect the observing strategy of WISE, 
observations were assumed to happen in a narrow range of solar elongation about 90$^\circ$.  
We then computed the average detection probability ${\cal P}(D,p_{\rm V})$ and analyzed it
as a function of $p_{\rm V}$. This test shows that the detection probability is relatively insensitive 
to asteroid albedo, at least in the size range of NEOs detected by NEOWISE ($D \gtrsim 100$ m).  
For example, for $D=0.3$ km, the detection probability decreases from $\simeq 4.2$\% for $p_{\rm V}=0$ 
to $\simeq 3.1$\% for $p_{\rm V}=0.5$. 

\subsection{Albedo distribution} 
There are three model parameters related to the albedo distribution. Following Wright et al. (2016),
we assume that the differential albedo distribution of NEOs can be approximated by a sum of two Rayleigh distributions
\begin{equation}
  p(p_{\rm V}) =
    f_{\rm d} {p_{\rm V} \over d^2 } \exp \left(-{p_{\rm V}^2 \over 2d^2}\right) + (1-f_{\rm d}) 
{ p_{\rm V} \over b^2 } \exp \left(-{p_{\rm V}^2 \over 2b^2}\right) \ , 
\label{wright}
\end{equation}          
with parameters $f_{\rm d}$, $d$ (the scale parameter for low-albedo or $d$ark NEOs) and $b$ (the scale parameter 
for high-albedo or $b$right NEOs), 
where $f_{\rm d}$ is the fraction of NEOs in the low-albedo Rayleigh distribution (the first term in Eq. (\ref{wright})). 
This functional form has fewer parameters than the double Gaussian distribution in 
Mainzer et al. (2011) and $p(p_{\rm V})$ falls to zero for $p_{\rm V} \rightarrow 0$ -- 
a desirable property of any physical model. 

Wright et al. (2016) determined $f_{\rm d}=0.253$, $d=0.030$ and $b=0.168$ for NEOs detected by cryogenic 
NEOWISE. Here we assume that Eq. (\ref{wright}) can be used for the debiased population as well and determine 
$f_{\rm d}$, $d$ and $b$ via the \texttt{MultiNest} fit. Mainzer et al. (2011) did not find any strong 
evidence for a correlation between albedo and size. For simplicity, we can thus assume that $f_{\rm d}$, 
$d$ and $b$ are unchanging with size (Sect 3.5).

\subsection{Combining CSS and NEOWISE}
 
The main objective of our work is to calibrate NEOMOD3 simultaneously from the CSS and NEOWISE data. 
CSS has a large number of detections, over $15,\!000$ NEOs from 2013 to 2022, which helps to accurately 
characterize  the absolute magnitude distribution of NEOs as faint as $H=28$. The NEOWISE data set 
gives us the albedo distribution of NEOs and allows us to convert the absolute magnitude 
distribution into the size distribution (Sect. 3.8).

In Paper I, we described a method that can be used 
to combine constraints from any number of surveys, and illustrated it for the 703 and G96 telescopes. 
In Paper II, we used the same method to combine the G96 data from 2013--2016 (before 
the G96 camera upgrade) with the G96 data from 2016--2022 (after the G96 camera upgrade). The method
consists in dealing with the surveys separately and evaluating the likelihood term in Eq. (\ref{like})
for each of them. The likelihood terms of different surveys are then simply summed up. We 
previously developed and used this method for visible surveys but it can be used for infrared surveys 
as well.  

To use this method here, we would need to compute the detection probability of NEOWISE as a function 
of the orbital elements $a,e,i$ (orbital longitudes can be ignored in the first approximation, but
see JeongAhn \& Malhotra 2014), absolute magnitude $H$ (or 
diameter $D$) and visible albedo $p_{\rm V}$. The detection probability has two parts: the geometric 
detection probability that an object will appear in WISE images and the photometric detection probability. 
The photometric detection probability is obtained from Eq. (\ref{w3}). To evaluate the geometric 
probability, we would need to collect the pointing history of WISE and link it with the Asteroid Survey 
Simulator (AstSim) package (Naidu et al. 2017), in much the same way this was done for CSS (Papers I 
and II). There would be no convenient way around this if the WISE observations were used on their own. Here, however,
CSS provides a much stronger constraint on the absolute magnitude distribution. In this situation, 
it makes better sense to fix parameters of the base model from CSS (Paper II) and infer the (debiased) 
albedo distribution from NEOWISE.

\subsection{Simple \texttt{MultiNest} fits}

A simple (biased) visible-albedo model of the NEO population can be defined as 
\begin{equation}
{\cal M}_{\rm b}(p_{\rm V};a,e,i,D) =  {\cal P}(p_V;a,e,i,D)\, p(p_{\rm V})\ ,
\label{bmodel}
\end{equation}
where ${\cal P}$ is the NEOWISE (photometric) detection probability (Sect. 3.2), and $p(p_{\rm V})$, as given 
in Eq. (\ref{wright}), is assumed to be independent of $a,e,i,D$. We consider 50 albedo bins for $0<p_{\rm V}<1$ 
and produce a binned version of ${\cal M}_{\rm b}$ (with the standard binning in $a,e,i$; Paper I). We 
only consider bins in $a,e,i,D$ where there were NEOWISE detections -- all other bins are ignored. 
For each detected object, we find the bin in $(a,e,i,D)$ to which it belongs, and compute the detection 
probability ${\cal P}$ for fixed $a,e,i,D$ and changing $p_{\rm V}$. This is done by placing a large 
number of bodies in each albedo bin, adopting the same diameter for all of them from NEOWISE,
running the NEATM model for all of them to determine the $W3$ magnitude in each case, 
and averaging the detection probability in the $W3$ band (Eq. \ref{w3}) over the 
whole sample (Sect. 3.2). 

In a bin $k$ in $(a,e,i,D)$, where there were $n_k=\sum_l n_{k,l}>0$ NEOWISE detections (typically $n_k=1$), 
where index $l$ runs over the albedo bins, we define $\lambda_{k,l} = {\cal M}_{\rm b}(p_{\rm V};a,e,i,D)$
and normalize it such that $\sum_l \lambda_{k,l} = n_k$ (we are not interested in the absolute calibration). 
The log-likelihood in \texttt{MultiNest} is defined as 
\begin{equation}
{\cal L} = - \sum_{k,l} \lambda_{k,l} + \sum_{k,l} n_{k,l} \ln \lambda_{k,l} \ ,
\label{like2}
\end{equation}
where index $k$ runs over bins in $a,e,i,D$ with NEOWISE detections (and index $l$ over all albedo bins).
\texttt{MultiNest} is then asked to determine parameters $f_{\rm d}$, $d$ and $b$ (Eq. \ref{wright}) by 
maximizing the log-likelihood in Eq. (\ref{like2}). This gives us, via Eq. (\ref{wright}), the intrinsic (debiased)
albedo distribution of NEOs. Note that the simple albedo model, as described here, does not need any input from NEOMOD.

\subsection{Complex \texttt{MultiNest} fits}

The simple albedo model can be generalized to account for the fact that different NEO sources may have 
different contribution to NEOs with low and high albedos (Morbidelli et al. 2020). This is done by 
generalizing $f_{\rm d}$ to have $n_{\rm s}=12$ coefficients $f_{{\rm d},j}$ that define the contribution of 
dark NEOs (i.e., NEOs in the low-albedo Rayleigh distribution in Eq. (\ref{wright})) individually for 
each source. In this case, the biased model is defined as 
\begin{equation}
{\cal M}_{\rm b}(p_{\rm V};a,e,i,D) = {\cal P}(p_{\rm V};a,e,i,D) \sum_{j=1}^{n_s} p_j(p_{\rm V})\, 
\alpha_j(H(D,p_{\rm V}))\,  p_{q^*,j}(a,e,i,H(D,p_{\rm V}))\,
\label{cmodel}
\end{equation}   
with 
\begin{equation}
p_j(p_{\rm V}) = f_{{\rm d},j} {p_{\rm V} \over d^2 } \exp \left(-{p_{\rm V}^2 \over 2d^2}\right) + (1-f_{{\rm d},j}) 
{ p_{\rm V} \over b^2 } \exp \left(-{p_{\rm V}^2 \over 2b^2}\right) \ , 
\label{wright2}
\end{equation}          
being the albedo distribution of source $j$. Here, $H(D,p_{\rm V})=-5\log_{10}(D\sqrt{p_{\rm V}}/c)$ with 
$c=1329$~km (Russel 1916). 
The contributions of different sources, $\alpha_j(H)$, and $p_{q^*,j}(a,e,i,H)$ are obtained 
from NEOMOD2 (these parameters are held fixed in the new fit). Again, as we are not interested in the absolute calibration, 
we define 
$\lambda_{k,l} = {\cal M}_{\rm b}(p_{\rm V};a,e,i,D)$ and normalize it such that $\sum_l \lambda_{k,l} = n_k$.
The \texttt{MultiNest} code is asked to determine the 14 parameters $f_{{\rm d},j}$, $d$ and $b$ by maximizing 
the log-likelihood in Eq. (\ref{like2}).\footnote{Note that this algorithm does not account for a viable 
possibility that the albedo distribution of NEOs from source $j$ can be size dependent (see Sect. 5.1).} 

\subsection{A note on coupling of model parameters}

The two algorithms described in Sect. 3.5 and 3.6 represent a good compromise between: (1) simplicity (i.e., 
number of model parameters; complicated albedo models cannot be robustly constrained from the NEOWISE data), 
(2) realism (e.g., we cannot ignore obvious biases; Sects. 3.2 and 5.1), and (3) CPU expense. 
We experimented with several different methods. 
For example, we explored algorithms to simultaneously determine the CSS and NEOWISE parameters in a 
single fit. For the complex \texttt{MultiNest} fit (Sect. 3.6), this represents 30 model parameters for 
CSS and 14 parameters for NEOWISE (twelve coefficients $f_{{\rm d},i}$, $d$ and $b$). In this case, we obtained 
the same values (and uncertainties) of model parameters as in the method described in Sect. 3.6. This shows 
that the CSS and NEOWISE parameters are uncorrelated. The 44 parameter approach is, however, very CPU expensive. 

\subsection{From $H$ and $p_{\rm V}$ to the size distribution}

It is not obvious how to convert the absolute magnitude and albedo distributions to the size distribution. 
This is because the albedo distribution, as obtained from NEOWISE,
\begin{equation}
  \phi(p_{\rm V}) = {{\rm d}N \over {\rm d}p_{\rm V}}\Bigr|_{D={\rm const}}\ ,
\label{phi1}
\end{equation}
is the albedo distribution of NEOs for a {\it fixed size} (or in a size range), and, for a simple conversion, we 
would need the albedo distribution for a {\it fixed absolute magnitude} (or in an absolute-magnitude range),
\begin{equation}
  \phi'(p_{\rm V}) = {{\rm d}N \over {\rm d}p_{\rm V}}\Bigr|_{H={\rm const}}\ .
\label{phi2}
\end{equation}
These two albedo distributions are different, $\phi(p_{\rm V}) \neq  \phi'(p_{\rm V})$,
because the distribution in the absolute-magnitude range has a larger contribution of asteroids with higher 
albedos (Appendix A).\footnote{Mainzer et al. (2011) faced 
the same problem and employed a Monte Carlo algorithm to obtain the size distribution. According to our tests, 
their algorithm is not rigorous and can lead to a factor of $\sim 2$ differences in the inferred size
distribution. This is because one cannot combine the {\it size-based} albedo distribution $\phi(p_{\rm V})$ and 
the absolute-magnitude distribution of NEOs to directly infer the size distribution. Instead, one has to 
resolve the inverse problem presented by Eq. (\ref{zeta}).}        

It can be shown that the three differential distributions, $n(H)={\rm d} N/{\rm d}H$, $\psi(D)={\rm d} 
N/{\rm d}D$ and $\phi(p_{\rm V})$ are related via the integral equation
\begin{equation}
n(H)={1 \over \kappa} \int_0^1  {\rm d} p_{\rm V}\, \phi(p_{\rm V})\, D \psi(D) \ ,
\label{zeta}
\end{equation} 
where $\kappa = -5/\ln 10$ and $D=D(H,p_{\rm V})=c 10^{-H/5}/\sqrt{p_{\rm V}}$ ($D$ must be substituted for $H$ and
$p_{\rm V}$ {\it before} the integral is evaluated). Eq. (\ref{zeta}) 
needs to be solved to obtain $\psi(D)$. We experimented with several approaches to this problem. It turns out 
that Eq. (\ref{zeta}) can be transformed, via substitutions of variables, to the integral Volterra equation of 
the first kind. It can be inverted to obtain $\psi(D)$ via the matrix inversion algorithm (Press et al. 1992) 
or Fourier transform (Muinonen et al. 1995).  

We opted for a different method in this work. We assumed that $\psi(D)$ can be parameterized 
by cubic splines in much the same way as $n(H)$ (Paper I), used the same number of segments for 
$\psi(D)$ as for $n(H)$ and converted the segment boundaries from $H$ to $D$ with a reference albedo 
$p_{\rm V,conv}$. The Simplex algorithm from Numerical Recipes (Press et al. 1992) was then employed 
to minimize a $\chi^2$-like quantity, and find the (cumulative) power-slope indices $\beta_j$ in all segments, 
and $p_{\rm V,conv}$. This procedure works perfectly well (Section 4). We tested it by first determining $\psi(D)$,
and then computing new $n(H)$ from $\psi(D)$ and $\phi(p_{\rm V})$ via Eq. (\ref{zeta}); this recovers 
the original distribution $n(H)$ without any significant errors. The spline approach described here has the 
advantage of having $\psi(D)$ immediately represented by splines -- the slopes in each segment have physical
meaning and the size distribution is easy to generate (e.g., in NEOMOD Simulator).

\section{Results}

\subsection{Simple fits}

We first discuss results from the simple \texttt{MultiNest} fits (Sect. 3.5). Table 1 reports the 
median and uncertainties for three model parameters that we obtain from a global fit to 
NEOWISE. Globally, the debiased albedo distribution of NEOs can be represented by Eq. (\ref{wright})
with $d=0.029\pm0.003$, $b=0.170 \pm 0.006$ and $f_{\rm d}=0.233 \pm 0.030$. This 
compares well with Wright et al. (2016), who found $d=0.030$, $b=0.167$ and $f_{\rm d}=0.253$ from a 
direct fit to the (biased) albedo distribution of NEOs detected by NEOWISE, and shows that the thermal 
infrared bias 
(Sect. 3.2) is not excessively important. Nominally, our best-fit $f_{\rm d}$ value is slightly lower 
than the one from Wright et al. (2016) (but note the large uncertainty), which means that the contribution 
of dark NEOs is slightly reduced in the debiased distribution, exactly as one would expect when 
the thermal bias is accounted for. 
Unfortunately, with the relatively small statistics from cryogenic NEOWISE detections, the 
uncertainties of the derived parameters are relatively large (Table 1 and Fig. \ref{triangle}).       
A comparison of the biased model with NEOWISE detections (Fig.~\ref{difalb}) demonstrates that
the model is acceptable. 

We used the method described in Sect. 3.8 to determine the size distribution of NEOs (Fig. \ref{sizedist}).
The best-fit size distribution is represented by splines in six diameter
segments (Table 2). We find a relatively steep slope for $D<50$ m ($\beta \simeq 2.5$--2.8) and 
a bending, concave profile for $D>100$ m. We estimate $\simeq 6.5\times10^6$ NEOs with $D>10$ m, 
$\simeq 30,\!000$ NEOs with $D>100$ m, and $\simeq 780$ NEOs with $D>1$ km. These estimates
were obtained from the global (simple) fit where the albedo distribution was held constant over the
whole range of diameters.

We also considered cases with the size-dependent albedo distribution, $\phi=\phi(p_{\rm V};D)$.
The motivation for this comes from the NEOWISE data. For example, the mean albedo of NEOs computed from all 
cryogenic NEOWISE measurements is $\langle p_{\rm V}\rangle=0.165$. If the NEOWISE detections are split 
according to object's size, however, we find that the mean albedo for $D>1$ km is $\langle p_{\rm V} \rangle 
= 0.135$ and the mean albedo for $D<1$ km is $\langle p_{\rm V} \rangle = 0.182$, suggesting some 
dependence of albedo on size.\footnote{This
  trend with smaller NEOs having (slightly) higher albedos is opposite to that expected from the
  thermal bias. It probably reflects the size-dependent contribution of main belt sources
  to NEOs (Sect. 4.2).}
To test the possible size dependence, simple \texttt{MultiNest} fits
were performed for NEOs of different sizes. We found, indeed, that the parameters $f_{\rm d}$, $d$ and $b$
change with size (Fig. \ref{linear}).\footnote{The formal uncertainty of $f_{\rm d}$ is large and the 
extrapolation to $D<100$ m is even more uncertain.} The results of these fits were interpolated to obtain
$\phi(p_{\rm V};D)$. The size distribution was then constructed with $\phi(p_{\rm V};D)$ (Fig. \ref{sizedist1}).
Table 3 reports our best estimates for the number of NEOs for the size-independent and
size-dependent albedo distributions. 
 
Our estimates are subject to several uncertainties: (1) We used the absolute-magnitude distribution
from NEOMOD2 where the dominant source of error -- at least for $H<25$ (CSS debiasing may have
introduced additional errors for $H>25$) -- was statistical in nature. In Paper II we estimated that
this represented the relative uncertainty of $\simeq3$\% for $H<25$. (2) There is an important
and potentially systematic uncertainty related to the absolute magnitude values reported
in the Minor Planet Center (MPC) catalog (Pravec et al. 2012, Harris \& Chodas 2023).
As we discussed in Paper II, due to shifting magnitude
values, the number of known NEOs with $H<17.75$ reported by MPC {\it decreased} by 49 from October 19, 2022
(our MPC download for Paper II) and March 13, 2023 (MPC download from Harris \& Chodas 2023).
If this trend holds, the number of $D>1$ km NEOs would be substantially revised. (3) Finally, there
is the uncertainty arising from the albedo distribution of NEOs. From the simple \texttt{MultiNest}
fits reported here, we conservatively estimate that the associated uncertainty is $<10$\%
for $D>100$ m (Table 3).\footnote{The uncertainty for $D<100$ m is larger because NEOWISE detected
  only a small number of NEOs with $D<100$ m. The albedo distribution of NEOs with $D<100$ m is therefore
  uncertain.} 

{\bf Accounting for items (1) and (3), we estimate $830 \pm 60$ NEOs with $D>1$ km and
$20,\!000 \pm 2,\!000$ NEOs with $D>140$ m. These are the values quoted in the abstract and 
conclusions. The ranges given here contain all estimates from different models reported in Table 3, 
and include the complex model results described in Sect. 4.2. This is a conservative approach, 
because the differences between different model results are generally larger than statistical 
uncertainties of individual models. See Paper II for a method that can be used to rescale these 
estimates from item (2).} 

Related to the NASA goal to discover 90\% of $D>140$ m NEOs, Fig. \ref{140} shows the absolute 
magnitude distribution for $D>140$ m NEOs. We used the NEOMOD Simulator and generated all model NEOs
with $D>140$ m. The results are plotted as a cumulative distribution of $H$ in Fig. \ref{140}. 
The distribution can be understood to indicate the fraction of $D>140$ m NEOs having magnitudes brighter 
than $H$. This information is relevant for the future telescopic surveys such as the Legacy Survey of 
Space and Time (LSST) of the Vera C. Rubin Observatory. For example, to reach a 90\% completion 
for $D>140$ m, telescopic observations would need to detect all NEOs brighter than $H=22.2$ or
$>90$\% of NEOs brighter than $H=24$. For reference, the current completness for $H<22$ and $H<24$ 
is only $\simeq 35$\% and $\simeq 10$\%, respectively (Paper II).

\subsection{Complex fits}

While the simple models described in the previous section can be used to infer the albedo dependence on
size, they cannot account for any albedo variation with orbit. There is some evidence in the
NEOWISE data that the albedo distribution can be orbit dependent. For example, NEOs with $D>1$ km and
$p_{\rm V}<0.1$ represent only $\simeq40$\% of all NEOs with $D>1$ km for $a<2$ au, but $\simeq56$\%
for $a>2$ au, suggesting that the fraction of dark NEOs increases with the semimajor axis
(Fig. \ref{baloons}).
This trend is expected because NEOs should reflect the taxonomic distribution of asteroids 
in the main belt, where dark (C-complex) bodies become more common with increasing semimajor
axis (DeMeo et al. 2009, Mainzer et al. 2019, Marsset et al. 2022). This motivates us to consider the complex
\texttt{MultiNest} fits from Sect. 3.6, where individual main belt sources can have different
contributions to dark and bright NEOs. 

Table 4 and Fig. \ref{triangle2} report model parameters from the complex \texttt{MultiNest}
fit. The complex model matches the NEOWISE data better than our simple model. The statistical
preference for a model is given by the Bayes factor evaluated by \texttt{MultiNest}. We obtain
$\Delta \ln {\cal Z} = 15.9$, indicating a strong preference for the complex model.
This can be readily understood because the complex model correctly emulates both the albedo
dependence on size (Fig. \ref{alb3}) {\it and} orbit (Fig. \ref{alb5}). The fraction of dark NEOs
($p_{\rm V}<0.1$) is found to increase with the semimajor axis. 
This is expected because dark (C-complex) asteroids are more common near NEO sources
in the outer main belt. For $1<a<2$ au the fraction of NEOs with $p_{\rm V}<0.1$ is $25\pm15$\%;
it increases to $65\pm15$\% at $a \sim 3$ au.\footnote{The relative paucity of dark NEOs
  detected by NEOWISE for $a<1$ au (or $q<0.25$ au) has been suggested 
  to result from catastrophic disruptions of dark, primitive, and presumably fragile NEOs that
  evolve onto orbits with low perihelion distances (Morbidelli et al. 2020). This effect was
  included in NEOMOD2 but we did not distinguish between bright and dark NEOs in Paper II.}  

The NEOWISE data do not provide sufficint information to constrain all (complex) model parameters. 
For example, the posterior distribution for $f_{{\rm d},j}$ parameters corresponding
to 7:3, 9:4 and comets is nearly uniform between 0 and 1 (Fig. \ref{triangle2}). This happens 
because these sources do not have a significant contribution to NEOs anyway (NEOMOD2 only gives 
a $< 2$\% contribution for them; Paper II). In some cases, such as Hungarias, we only obtain an upper 
bound with $f_{\rm d}<22$\%. In other cases, such as the 11:5 resonance, we obtain a 
lower bound with $f_{\rm d}>72$\%. The upper (lower) limits mean that the low-albedo (high-albedo) 
bodies should represent the great majority of NEOs produced from that source.

In general, the contribution of sources to dark NEOs correlates with the semimajor axis.
The inner belt sources such as $\nu_6$ and 3:1 have low contributions, and the outer belt sources such as 
11:5 and 2:1 have high contributions (Fig. \ref{type}). A similar trend was reported in Morbidelli et al. (2020).
As in Morbidelli et al. (2020), here we also find a relatively large contribution to dark NEOs
from Phocaeas ($\simeq 50$\% for $p_{\rm V}<0.1$ in Morbidelli et al. and $>65$\% for $p_{\rm V}<0.1$
here).\footnote{To compute the fraction of $p_{\rm V}<0.1$ NEOs from Phocaeas, we used $f_{\rm d}({\rm Pho})=0.76$ 
(Table 3) and summed up the contributions of dark and bright Rayleigh distributions from 
Phocaeas to $p_{\rm V}<0.1$.} This is inconsistent with other observational evidence which suggests that 
Phocaeas are mostly bright (S-type) asteroids (DeMeo et al. 2009; about 1/3 of Phocaeas have 
$p_{\rm V}<0.1$, Mainzer et al. 2019).\footnote{Novakovi\'c
  et al. (2017) identified a dark and relatively young asteroid 
family in the Phocaea region (the Tamara family; age $264\pm43$ Myr). They estimated 
that $\sim 500$ of its members with $17<H<19.35$ reached the NEO orbits in total. With the mean 
lifetime of NEOs from the Phocaea source, 13.5 Myr from NEOMOD2, we can estimate that there
should be $\sim 37$ dark Tamara family NEOs in a steady state. For comparison, there are $\simeq 3,\!500$
NEOs with $H<19.35$ (Paper II), of which $\sim 0.09 \times 0.76$ should be dark Phocaeas
(according to the contribution of Phocaeas to large NEOs from Paper II, $\simeq 9$\%,
and the dark fraction found here, $\sim76$\%). This gives $\sim 240$, suggesting that the Tamara
family cannot be a major contributor.} The problem may arise from the relatively low statistics of 
NEOWISE detections: a handful of dark NEOs were detected by NEOWISE on high-inclination orbits where 
the Phocaea source is expected to contribute. Either that, or we are missing a source of dark NEOs 
on high inclination orbits.  
 
Figure \ref{dalb} illustrates the size-dependent albedo distribution of NEOs from the {\it debiased}
complex model. The distributions shown here for $D>0.1$ km are in good agreement with those
obtained with different size cuts in the simple model (Sect. 4.1). For $1<D<3$ km,
the complex model indicates $f_{\rm d}=0.33$, in a close match to the result reported in Fig.
\ref{linear}. For $0.1<D<0.3$ km, we have $f_{\rm d}=0.23$, slightly higher than $f_{\rm d}=0.18$ from
the simple model. Some differences are expected given the different schemes employed in our simple
and complex models.\footnote{The simple model is firmly tied to NEOWISE and gives us the albedo
distribution for orbits of NEOs detected by NEOWISE, whereas the complex model weights albedos with
the help of the orbital distribution from NEOMOD2 (Fig. \ref{unb}; Sect. 5.1). The slightly
lower $f_{\rm d}$ values obtained from the simple model presumably reflect the orbital bias (see 
Sect. 5.1).} The albedo distribution
for $D<0.1$ km is an extrapolation with NEOMOD2 and the complex model parameters listed in Table 3.
If these results are correct, the importance of the dark Rayleigh peak continues to diminish for
$D<0.1$ km, indicating that (very) small NEOs are on average (much) brighter than large NEOs.  

The complex model inferences for the size distribution of NEOs are consistent with those obtained 
from the simple model. Because the $f_{\rm d}$ values tend to be slightly larger in the complex model, 
here we obtain slightly higher population estimates than $N_3(D)$ reported in Table 3, nominally 
873 NEOs with $D>1$ km and 19,500 NEOs with $D>140$ m. This is well within the range of uncertainties 
discussed in Section 4.1. The population estimates from the complex model could be favored over those
obtained in the simple model, because the complex model is more successful in reproducing various orbital
dependences. In some cases, however, such as the Phocaea case discussed above (also see Morbidelli et 
al. 2020), the inferences obtained from the complex model are somewhat uncertain. In this situation, 
we prefer to report the full range of population estimates from the simple and complex models. This 
is why the abstract and conclusions give $830 \pm 60$ NEOs with $D>1$ km and $20,\!000 \pm 2,\!000$ 
NEOs with $D>140$ m.  

The reference albedo value $p_{\rm V,ref}$ for an approximate conversion of the absolute magnitude 
distribution to the size distribution (e.g., Harris \& Chodas 2021) is a function of absolute magnitude. 
We recommend $p_{\rm V,ref}\simeq0.15$ for $H<18$, $p_{\rm V,ref}\simeq0.16$ for $18<H<22$, and 
$p_{\rm V,ref}\simeq0.18$ for $H>22$. 
  
\section{Discussion}

\subsection{Simple vs. complex model inferences}

There are at least two obvious biases in NEOWISE observations. The first one is the {\it thermal} 
infrared bias discussed in Section 3.2 (objects with low visual albedo emit more thermal radiation and are more 
easily detected in infrared wavelengths). Our simple model rigorously accounts for the thermal bias
(Sect. 3.5). The second one is the {\it orbital} bias: the NEOWISE data set is biased toward detection 
of NEOs with small heliocentric distances. These NEOs are warmer and more easily 
detected in thermal infrared. We know that NEOs at small heliocentric distances predominantly sample sources 
in the inner asteroid belt; they are more likely to have higher albedos than NEOs on larger orbits. This 
means that NEOWISE is biased toward higher albedos. This is not something we can account for in the simple 
model. The simple model calibrates the albedo distribution on NEOs detected by NEOWISE (the thermal bias 
is accounted for) and adopts it for NEOs in general. The simple model should thus be biased toward 
higher albedos as well (due to the orbital bias).

The debiased albedo distribution obtained from the complex model does not suffer from this
limitation, at least not as much as the simple model, because it adopts the orbital distribution
of NEOs from NEOMOD2. For example, the $\nu_6$ resonance produces evolved NEOs with $a<2$ au. These 
bodies escape from the inner asteroid belt and often have $p_V>0.1$. The $\nu_6$ resonance is thus 
assigned a relatively low value of the parameter $f_{\rm d}$, and the albedo distribution -- specific 
for the $\nu_6$ resonance -- is then extended with a proper weight to the whole NEO population.

The same applies to other sources as well. So, at least in principle, the complex model should
give us a more realistic albedo distribution of NEOs, including its proper scaling with size and orbit.
This may explain some of the differences discussed in Sect. 4.2. Note that these differences are
not large, however, suggesting that the orbital bias in the simple model is not overwhelmingly 
important. We discuss the simple model in this work because the simple model is firmly tied to 
NEOWISE observations, does not require additional assumptions (e.g., related to how NEOs sample 
various main belt sources), and allows us to test the albedo dependence on size. The fact that 
the simple and complex models lead to consistent results is reassuring.    

Additional uncertainties arise because even the complex model does not account for the possibility
that the albedo distribution of NEOs from source $j$ can be size-dependent (e.g., because
the low- and high-albedo main-belt asteroids near that source have different size distributions).
The model defines the albedo distribution from source $j$ as unchanging with size, and injects the
size and orbit dependence of NEO albedo via the size-dependent contribution of sources,
$\alpha_j(H)$ (Paper II). Investigations into more complete albedo models are left for future work.

\subsection{Relationship to main belt asteroids}

Some features of the complex model seem surprising. For example, according to
Fig. \ref{type}, the $\nu_6$ resonance is inferred to produce only $\simeq 20$\% of NEOs with
$p_{\rm V}<0.1$. If we look in the immediate neighborhood of the $\nu_6$ resonance in the main belt,
we find that $\simeq 40$\% of asteroids with $1<D<2$ km have $p_{\rm V}<0.1$ (Mainzer et al.
2019). This can mean one of several things. In NEOMOD2, the $\nu_6$ source does not have much
contribution to NEOs with $D>1$ km (Paper II). The albedo distribution of $\nu_6$ is thus mainly
calibrated on small, sub-km NEOs detected by NEOWISE. Since these small bodies were not detected
by NEOWISE in the main belt, however, we cannot be sure that there really is a problem. A
similar argument applies to the 3:1 resonance as well. 

In more general terms, we find here that dark NEOs with $p_{\rm V}<0.1$ represent $\simeq 40$\%
of the NEO population (for $D \sim 1$ km). This is lower than the share of dark asteroids in
the main belt ($\sim 60$\% overall from WISE; Mainzer et al. 2019). The difference is in part caused by how
NEOs sample the main belt -- they preferentially come from the inner part of the belt 
where dark asteroids are less common. Overall, dark bodies with $p_{\rm V}<0.1$ contribute
to $\sim 40$\% of asteroids in the inner belt (2--2.5 au). The NEOWISE data also indicate that the albedo
distribution of inner belt asteroids may be size dependent. For example, dark bodies with $p_{\rm V}<0.1$
represent 55\% of inner belt asteroids with $D>10$ km, but only 27\% of inner belt asteroids
with $1<D<2$ km. 

\subsection{NEO population estimates}

We estimate $830\pm60$ NEOs with diameters $D>1$ km and $30,\!000 \pm 3,\!000$ NEOs with $D>100$ m
(Table 3). This can be compared to $\simeq 20,\!500 \pm 3000$ NEOs with $D>100$ m and $981 \pm 19$ NEOs with
$D>1$ km reported in Mainzer et al. (2011), and $\sim 1000$ NEOs with $D>1$ km in Morbidelli
et al. (2020). Our population estimate for $D>100$ m is $\sim 50$\% higher. We believe that our method better
approximates the debiased size distribution for $D<1$ km.  Our estimate for $D>1$ km is $\sim 15$\% lower.
We think that this happens because Mainzer et al. (2011) used an approximate Monte Carlo method to
infer the number of large NEOs. Here we infer it by inverting Eq. (\ref{zeta}), which is a more 
rigorous approach. 

The error estimates reported here combine various uncertainties related to our
inferences about the absolute magnitude distribution from NEOMOD2 and the albedo distribution from NEOWISE.
We find that the dominant source of error - at least
the one that we are able to characterize at the present time - reflects uncertainties in
the albedo distribution of NEOs. As we varied the debiasing
method and tweaked parameters in the \texttt{MultiNest} fits, we found that the estimates vary
by $\lesssim 10$\%. Hence our conservative error estimates, but note that systematic changes of
MPC magnitudes are not accounted for here (see Sect 4.1).

\subsection{Impact flux on the Earth}

Here we estimate the impact flux of NEOs on the Earth. This is done by combining the
absolute magnitude distribution from NEOMOD2, the albedo distribution from NEOWISE, and the intrinsic
impact probability, $P_{\rm i}(H)$, for NEO impacts on the Earth from Paper II.\footnote{The intrinsic 
impact probability $P_{\rm i}(H)$ is defined as the probability for one object in the NEO population 
with absolute magnitude $H$ to impact on the Earth in Myr.} The impact flux
is obtained by inverting Eq. (\ref{zeta}), where instead of $\psi(D)$ in the integrand there is
$P_{\rm i}(H) \psi(D)$. For reference, $P_{\rm i}=1.53\times10^{-3}$ Myr$^{-1}$ for $H=15$,
$P_{\rm i}=2.1\times10^{-3}$ Myr$^{-1}$ for $H=20$ and $P_{\rm i}=2.6\times10^{-3}$ Myr$^{-1}$ for $H=25$ 
(Paper II). Figure \ref{impact} shows the impact flux for the size-dependent albedo model, including 
the tidal disruption model from Paper II.\footnote{Tidal disruptions affect the impact profile for $D<100$ m. 
Without tidal disruption, the (cumulative) power-slope index for impacts of $D<100$ m NEOs is $\simeq 2.6$.
With tidal disruption, it steepens to $\simeq 3.1$.} Table 5 reports the number of impacts for several 
reference impactor diameters.

We estimate 1.51--1.74 impacts/Myr of $D>1$ km NEOs on the Earth. The average interval between impacts of
$D>1$ km is 570--660 kyr. This is shorter than the estimate given in Morbidelli et al. (2020)
who found the average interval $\simeq 750$ kyr. The difference reflects different population 
estimates and different impact probabilities adopted in these works. 
For $D>140$ m, we find 42--52 impacts/Myr and the average
interval between impacts 19--24 kyr. We can also compare our results with Nesvorn\'y et al. (2021),
where a different method was used for very large NEOs. They inferred 16--32 impacts/Gyr of $D>5$ km
NEOs on the Earth. Here we find $\sim 30$ such impacts (Fig. \ref{impact}), a value near the upper
end of the range given in Nesvorn\'y et al. (2021). The trend pointed out here, with the larger share
of dark bodies among large NEOs is consistent with Nesvorn\'y et al. (2021), who argued that dark
(primitive) asteroids represent about a half of very large impactors ($D\gtrsim5$ km) on the Earth.

For the smallest impactors shown in Fig. \ref{impact}, we find that the mean interval between impacts
of $D>10$ m NEOs is $\sim 40$ years. This is consistent with the results reported in Paper II (see 
the black solid line in Fig. \ref{impact}) given that the results presented here suggests that the 
albedo of small NEOs should be relatively high - effective $p_{\rm V}\simeq 0.18$ (instead of the usual 
reference $p_{\rm V}=0.14$, Paper II). This is a consequence of the $\nu_6$ resonance having a relatively 
large contribution for small and bright NEOs. The impact flux obtained here is a factor of $\sim 4$ 
below the impact flux estimate obtained from bolide observations ($\sim 10$ yr interval between 
impacts of $D>10$ m NEOs; Brown et al. 2013), which is a problem. 

The visible albedos of $D\sim10$ m NEOs obtained in this work may be too high. The albedo distribution 
of small, $D<100$ m NEOs was obtained here by calibrating the model on relatively large NEOs ($D>100$ m) 
detected by NEOWISE. In the complex model, we assumed that the number ratio of dark over bright bodies, 
as calibrated for individual sources on $D>100$ m NEOs, does not change for $D<100$ m. 
This assumption may be incorrect. For example, the contribution of dark asteroid families close to the 
$\nu_6$ and/or 3:1 sources may be insignificant for $D>100$ m, but important for $D<100$ m. If so, 
this would effectively lower the reference albedo. Another possibility is that the tidal disruption of 
NEOs during close planetary encounters (Paper II) disproportionally affects dark NEOs, perhaps 
because they are weak, and creates an excess of small dark NEOs on orbits with high impact probabilities 
on the Earth (this effect is not taken into account in the present work). 

\subsection{Lunar production function}

The radiometric ages, crater counts and size distribution extrapolations are the basis of empirical
models for impact cratering in the inner solar system (see Ivanov et al. 2002 for a review). The standard 
approach to this problem is to conduct crater counts on different lunar terrains and patch them together to 
estimate the \textit{lunar production function} (LPF), defined as the number of craters larger than diameter 
$D_{\it crater}$ produced on 1 km$^2$ of the lunar surface in Gyr. Here we estimate 
the current-day LPF from the size distribution of NEOs (also see Marchi et al. 2009). 
The results shown in Fig. \ref{impact} 
are carried over to lunar impacts with the standard Earth-to-Moon ratio (${\cal R}=20$; Paper I). 
We adopt the crater scaling laws from Johnson et al. (2016), for which a $D=1$-km NEO impactor 
makes a $D_{\rm crater}\simeq20$-km lunar crater, and a $D \simeq 40$-m NEO impactor makes a $D_{\rm crater} 
\simeq 1$ km lunar crater (see Morbidelli et al. (2018) for a discussion). 

Figure \ref{neukum} compares our LPF with those inferred from the crater counts in Hartmann (1995) 
and Neukum et al. (2001). This is not a one-to-one comparison for several different reasons. For 
example, the lunar craters with $D_{\rm crater}<1$ km are often secondaries (i.e., craters formed by 
re-impacting material ejected from a primary crater; Bierhaus et al. 2018). The secondaries are
not accounted for in our model. Also, there are not enough large craters with $D_{\rm crater} \gtrsim 10$ km 
on the young lunar terrains -- the empirical LPF for $D_{\rm crater} \gtrsim 10$ km must therefore be 
inferred from old lunar terrains, but the old lunar terrains may have seen impactor populations 
other than modern NEOs (Nesvorn\'y et al. 2022, 2023b). 

With these caveats in mind, we find that our LPF is roughly intermediate between LPFs reported in 
Hartmann (1995) and Neukum et al. (2001) (Fig. \ref{neukum}). For $D_{\rm crater}<1$ km, the empirical 
LPFs are somewhat steeper than our LPF possibly due to the contribution of secondaries (secondary craters
tend to have steep size distributions; Bierhaus et al. 2018). For some reason, our LPF runs below 
that of Hartmann (1995), indicating a problem with the absolute calibration, but nicely reproduces the slope 
transition near $D_{\rm crater}=1.5$ km (steeper for $D_{\rm crater}<1.5$ km, shallower for $D_{\rm crater}>1.5$ km). 
Neukum's LPF shows a broader transition near $D_{\rm crater}=5$ km, but the shape of this transition may be affected 
by crater counts on very old lunar terrains.       
 
\section{Conclusions}

The main results of this work are summarized as follows:
\begin{description}
\item (1) We developed approximate methods to debias the albedo distribution of NEOs detected by NEOWISE.
  The debiased albedo distribution can be accurately described by a sum of two Rayleigh
  distributions representing NEOs with low ($p_{\rm V}\lesssim0.1$) and high albedos ($p_{\rm V}\gtrsim0.1$). 
\item (2) There is good evidence that the albedo distribution of NEOs is {\it size} and {\it orbit} dependent. Smaller
  NEOs tend to have higher albedos than large NEOs. NEOs with evolved orbits below 2 au tend to have higher
  albedos than NEOs beyond 2 au.
\item (3) The debiased albedo distribution and absolute magnitude distribution of NEOs from NEOMOD2 (Paper 2)
  were used to infer the size distribution of NEOs. We estimate $830\pm60$ NEOs with diameters $D>1$ km and
  $20,\!000 \pm 2,\!000$ NEOs with $D>140$ m. See the bold paragraph in Sect. 4.1 for how these 
  estimates and their uncertainities were synthesized from different models (the range contains estimates 
  from all models investigated here). 
\item (4) The reference albedo value $p_{\rm V,ref}$ for an approximate conversion of the absolute magnitude 
  distribution to the size distribution is a function of absolute magnitude. We recommend $p_{\rm V,ref}\simeq0.15$ 
  for $H<18$, $p_{\rm V,ref}\simeq0.16$ for $18<H<22$, and $p_{\rm V,ref}\simeq0.18$ for $H>22$. 
\item (5) The intrinsic impact probability from NEOMOD2 was combined with the population
  estimates obtained here to infer the impact rates of NEOs on the Earth. We estimate the average interval
  between impacts of $D>1$ km NEOs about 640 kyr, and the average interval between impacts of $D>140$ m
  NEOs about $20,\!000$ yr.
\item (6) We used the NEO model to estimate the production function of lunar craters. The lunar production function (LPF)
is found to have an inflection point for $D_{\rm crater} \simeq 1.5$ km, with the steeper slope for 
$D_{\rm crater} < 1.5$ km and shallower slope for $D_{\rm crater} > 1.5$ km. A similar slope transition
was inferred from the lunar crater counts in Hartmann (1995).     
\item (7) The upgraded model, NEOMOD3, is available via the NEOMOD Simulator -- a user-friendly code
  that can be used to generate samples (orbits, sizes and albedos of NEOs) from the
  model.\footnote{\url{https://www.boulder.swri.edu/\~{}davidn/NEOMOD\_Simulator} and GitHub.}  
\end{description}

\section{Appendix A: Albedo bias in visible surveys}

Assume, for example, a bimodal (differential) distribution of albedos, $\phi(p_{\rm V})={\rm d}N/{\rm d}p_{\rm V}$,
with $\phi_{\rm dark}(p_{\rm V}) = \delta(p_{\rm V}-d)$ for dark objects and $\phi_{\rm bright}(p_{\rm V}) = \delta(p_{\rm V}-b)$ 
for bright objects, where $\delta$ are delta functions, and $b$ and $d$ are some 
characteristic albedo values of dark and bright objects, respectively. For example, Wright et al. (2016) 
found that the albedo distribution of NEOs detected by NEOWISE
can be approximated by a sum of two Rayleigh distributions with 
the scale factors $d=0.03$ and $b=0.168$. Assume, in addition, that the size distributions of dark 
and bright objects, $\psi(D) = {\rm d}N/{\rm d}D$, can be approximated by the same power law slope,
$\psi_{\rm dark}(D) = f_{\rm d} N_0 D^{-\alpha}$ for dark and $\psi_{\rm bright}(D) = (1-f_{\rm d}) N_0 D^{-\alpha}$
for bright, where $f_{\rm d}$ is the share of dark objects in the population, and $\alpha$ is fixed. 

\begin{figure}
\epsscale{0.6}
\plotone{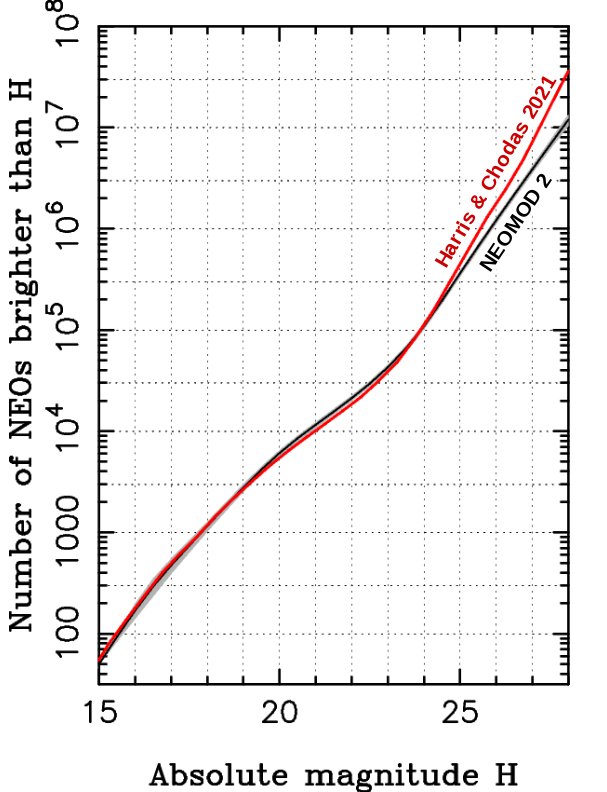}
\caption{The intrinsic (debiased) absolute magnitude distribution of NEOs from NEOMOD2
(Paper II, the black line is a median) is compared to the magnitude distribution from Harris \& Chodas (2021)
(red line). The gray area is the 3$\sigma$ envelope obtained from the posterior distribution 
computed by \texttt{MultiNest}. It contains -- by definition -- 99.7\% of our base model 
posteriors.}
\label{harris}
\end{figure}

\begin{figure}
\epsscale{0.8}
\plotone{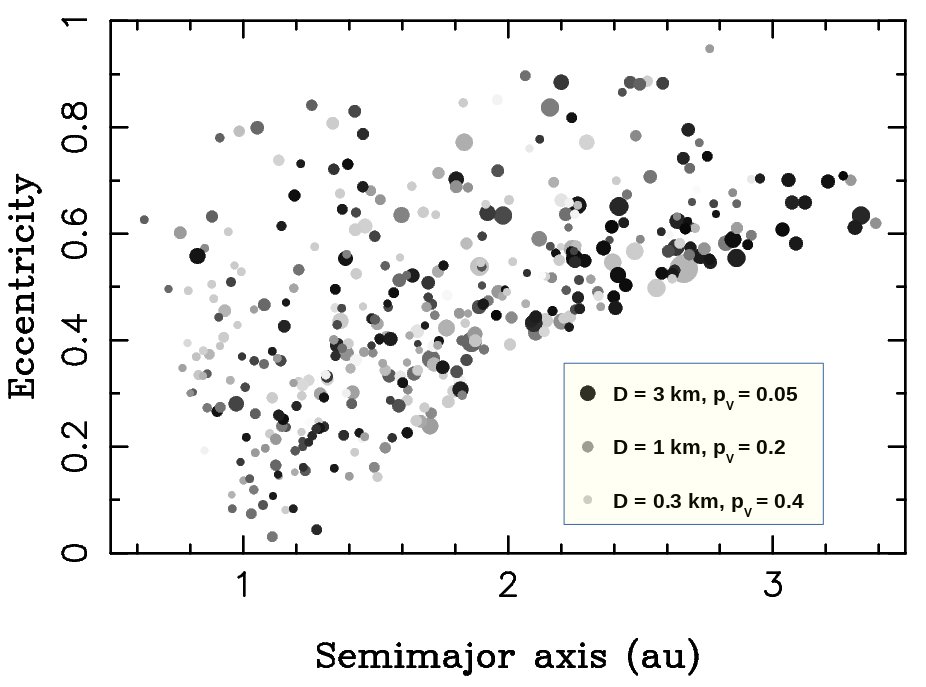}
\caption{The orbits and albedos of 428 NEOs detected during the cryogenic portion of the WISE mission. The darker 
the dot, the lower the albedo. The size of a dot is proportional to asteroid diameter. There is a general trend
with brighter asteroids being more prevalent for $a<2$ au (77\% of NEOWISE NEOs have $p_V>0.1$)
than for $a>2$ au (66\% have $p_V>0.1$).}
\label{baloons}
\end{figure}

\begin{figure}
\epsscale{0.7}
\plotone{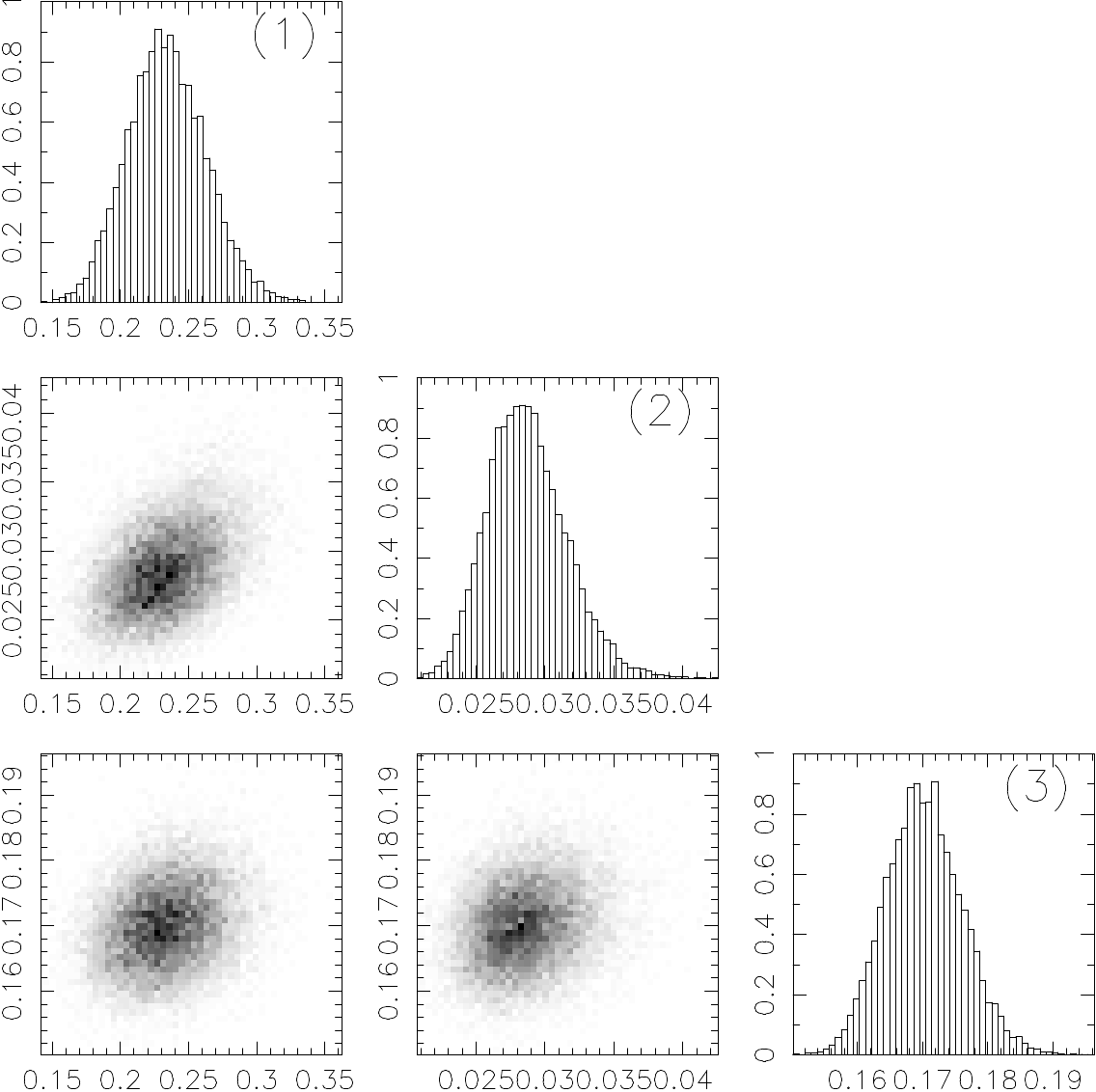}
\caption{The posterior distribution of model parameters from our simple \texttt{MultiNest} fit
(Sect.~3.5). The individual plots are labeled (1) to (3) following the model parameter sequence 
given in Table 1.}
\label{triangle}
\end{figure}

\begin{figure}
\epsscale{0.49}
\plotone{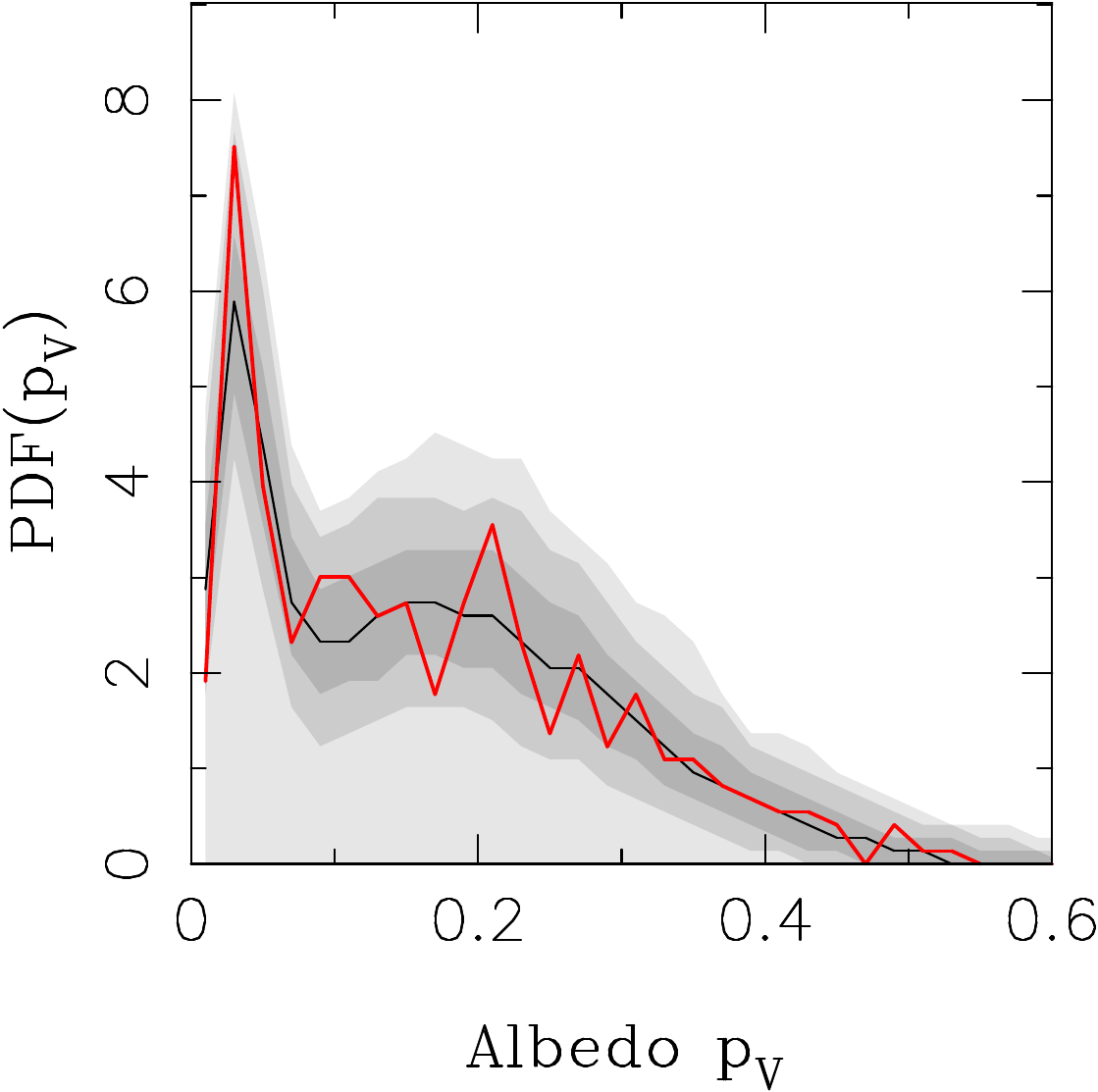}
\plotone{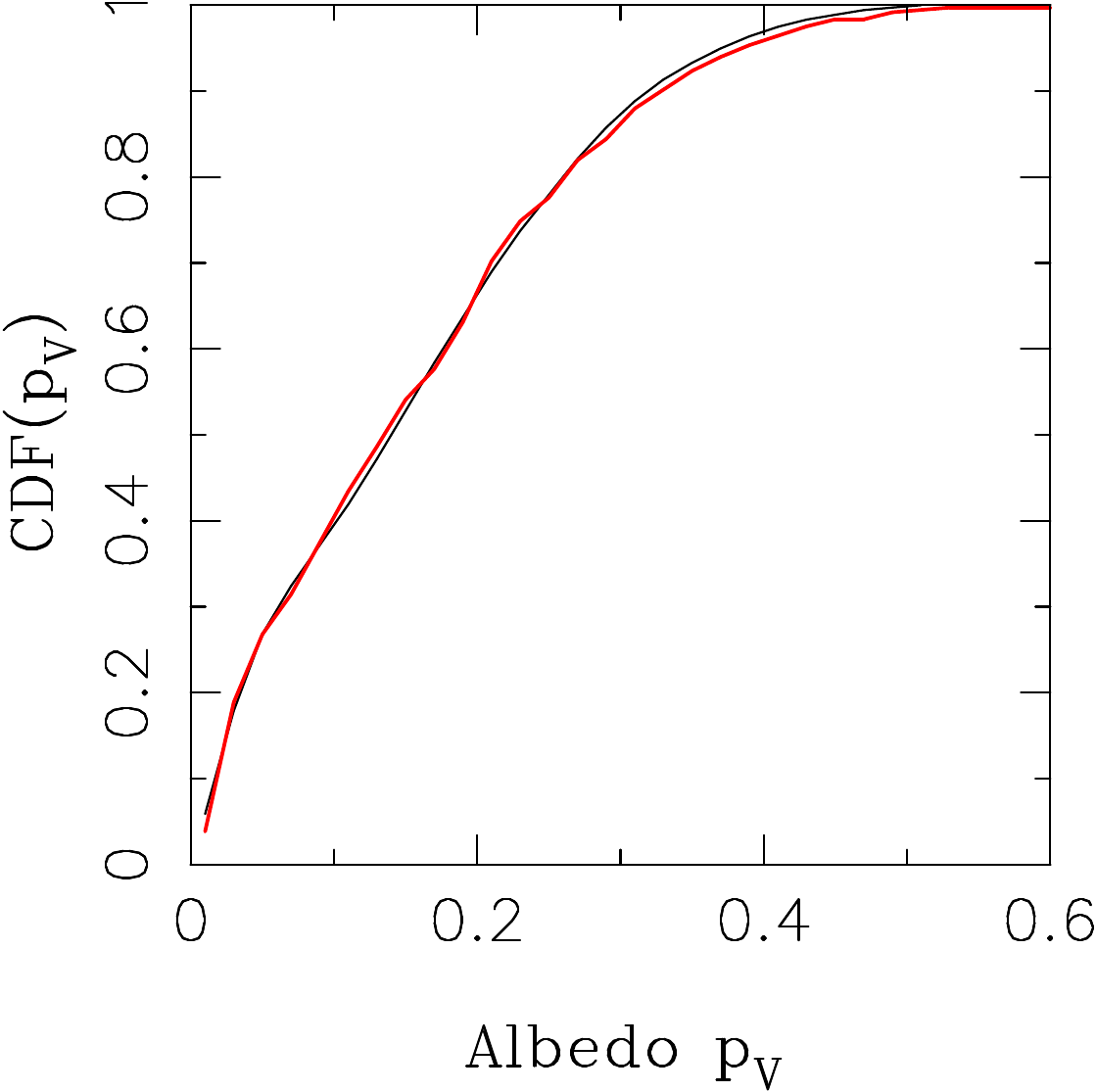}
\caption{The global model for the biased albedo distribution of NEOs (black lines are the median) is compared to
  NEOWISE detections (red lines). The plot on the left shows the differential distributions, the plot on the right shows the 
cumulative distributions. The shaded areas in the left panel are 1$\sigma$ (bold gray), 2$\sigma$ (medium) and 
3$\sigma$ (light gray) envelopes. We used the best-fit solution (i.e. the one with the maximum likelihood) 
from the base model and generated 10,000 random samples with 428 NEOs each (the sample size identical 
to the number of NEOs detected by NEOWISE in the model domain). The samples were biased and binned with the 
standard binning. We identified envelopes containing 68.3\% (1$\sigma$), 95.5\% (2$\sigma$) and 99.7\% 
(3$\sigma$) of samples and plotted them here. The Kolmogorov-Smirnov test indicates a large probability that
the two distributions -- the biased model and NEOWISE detections -- are the same.
}
\label{difalb}
\end{figure}

The (differential) magnitude distribution, $n(H)={\rm d} N/{\rm d}H$, can be obtained by evaluating the 
integral over all albedo values
\begin{equation}
n(H)={1 \over \kappa} \int_0^1  \, {\rm d} p_{\rm V} \, \phi(p_{\rm V}) D \psi(D) \ ,
\end{equation} 
where $\kappa = -5/\ln 10$, $D=D(H,p_{\rm V})= c 10^{-H/5}/\sqrt{p_{\rm V}}$, and $c=1329$ km. For the example 
discussed above, this gives $n(H) = N'_0 10^{\gamma H}$ with $\gamma = (\alpha-1)/5$, $N'_0=N_0 c^{-5\gamma} 
p_{\rm V,ref}^{5\gamma/2} / \kappa$, and the reference albedo $p_{\rm V,ref}^{5\gamma/2} = f_{\rm d} d^{5\gamma/2} + (1-f_{\rm b}) b^{5\gamma/2}$.
The reference albedo $p_{\rm V,ref}$ can be used to convert the absolute magnitude distribution to the size distribution. 
The real absolute magnitude distribution of NEOs is wavy (Figure \ref{harris}) with $\gamma \simeq 0.3$--0.55
(Papers I and II). For $\gamma = 0.4$, we have $5\gamma/2 = 1$ and the reference albedo is just a normal 
(weighted by $f_{\rm d}$) mean of $d$ and $b$. For the example from Wright et al. (2016), with 
$d=0.030$ and $b=0.168$, this gives $p_{\rm V,ref}=0.133$. For $\gamma=0.3$ and 0.5, we have $p_{\rm V,ref}=0.128$ and 
$p_{\rm V,ref}=0.137$, respectively.

Now, as for the albedo bias in a visual-magnitude limited survey, the sizes of the dark and bright objects with the 
same magnitude $H$ are $D_{\rm d}=c 10^{-H/5}/\sqrt{d}$ and $D_{\rm b}=c 10^{-H/5}/\sqrt{b}$. The fraction of dark objects
in a magnitude-limited survey is then $f'_{\rm d} = f_{\rm d} D_{\rm d}^{-\alpha} / [f_{\rm d} D_{\rm d}^{-\alpha} + 
(1-f_{\rm d}) D_b^{-\alpha}]$. This gives
\begin{equation}
f'_{\rm d} = f_{\rm d} \left[ f_{\rm d} + (1-f_{\rm d}) \left( {b \over d} \right)^{\alpha/2} \right]^{-1}\ .
\end{equation}   
For the example discussed above with $f_{\rm d}=0.253$ and $\gamma=0.4$, we have $\alpha=3$ and 
$f'_{\rm d}=0.025$. So, the bright objects would represent 97.5\% of all objects (even though their actual 
share in a size-limited sample is only 74.7\%). Additional complications would arise if the dark and
bright objects do not have the same power slope index or if the power slope index changes 
with size.         

\begin{figure}
\epsscale{0.6}
\plotone{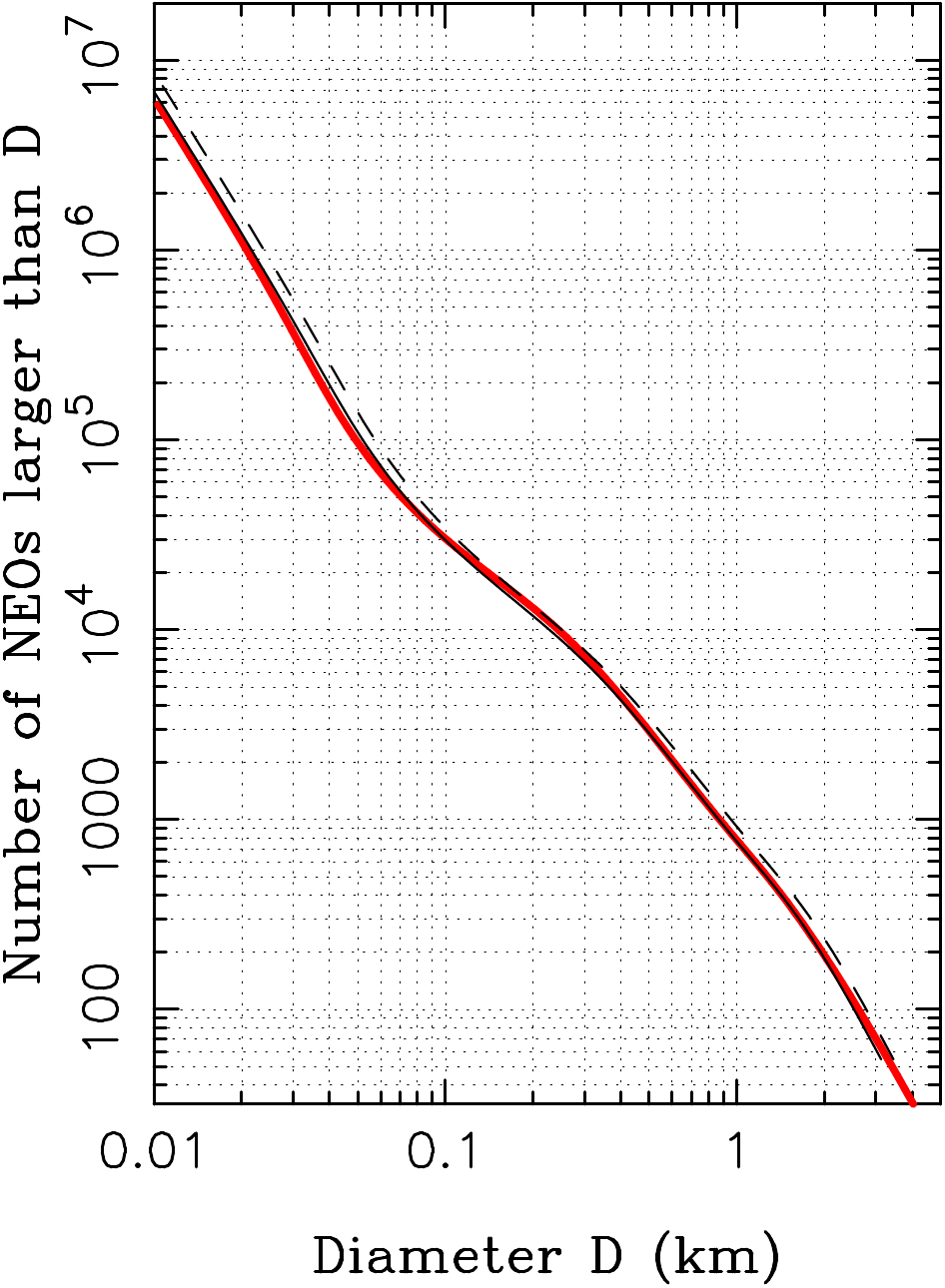}
\caption{The size distribution of NEOs from our simple model (red line; Sects. 3.5 and 3.8) is compared to 
the size distributions constructed from the best-fit absolute magnitude distribution in Paper II and 
reference visual albedo $p_{\rm V}=0.14$ (dashed line) and 0.17 (solid line). The albedo distribution
of NEOs used here comes from a global fit to the NEOWISE data. It is held constant over the whole
range of diameters shown in the plot.}
\label{sizedist}
\end{figure}

\begin{figure}
\epsscale{0.7}
\plotone{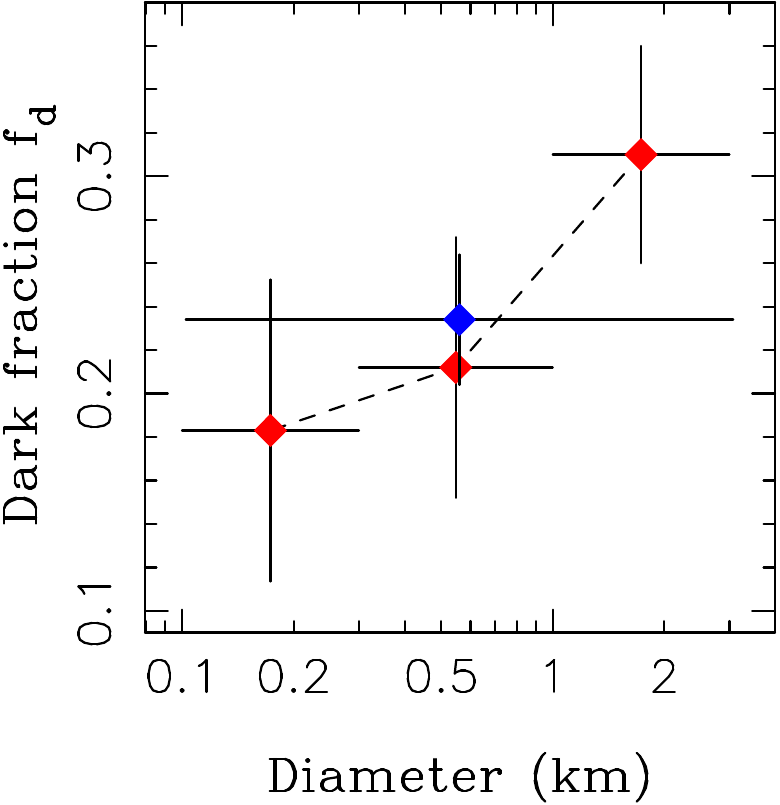}
\caption{The fraction of NEOs in the dark Rayleigh peak ($f_{\rm d}$; Sect. 3.3) obtained from
  different \texttt{MultiNest} fits. The blue symbol and the errors bars show results from
  the global (simple) fit to all NEOWISE observations. The red symbols and the errors bars show results
  for fits in different size ranges. The dashed lines indicate the interpolated values used in the model
  with the size-dependent albedo distribution.}
\label{linear}
\end{figure}

\begin{figure}
\epsscale{0.6}
\plotone{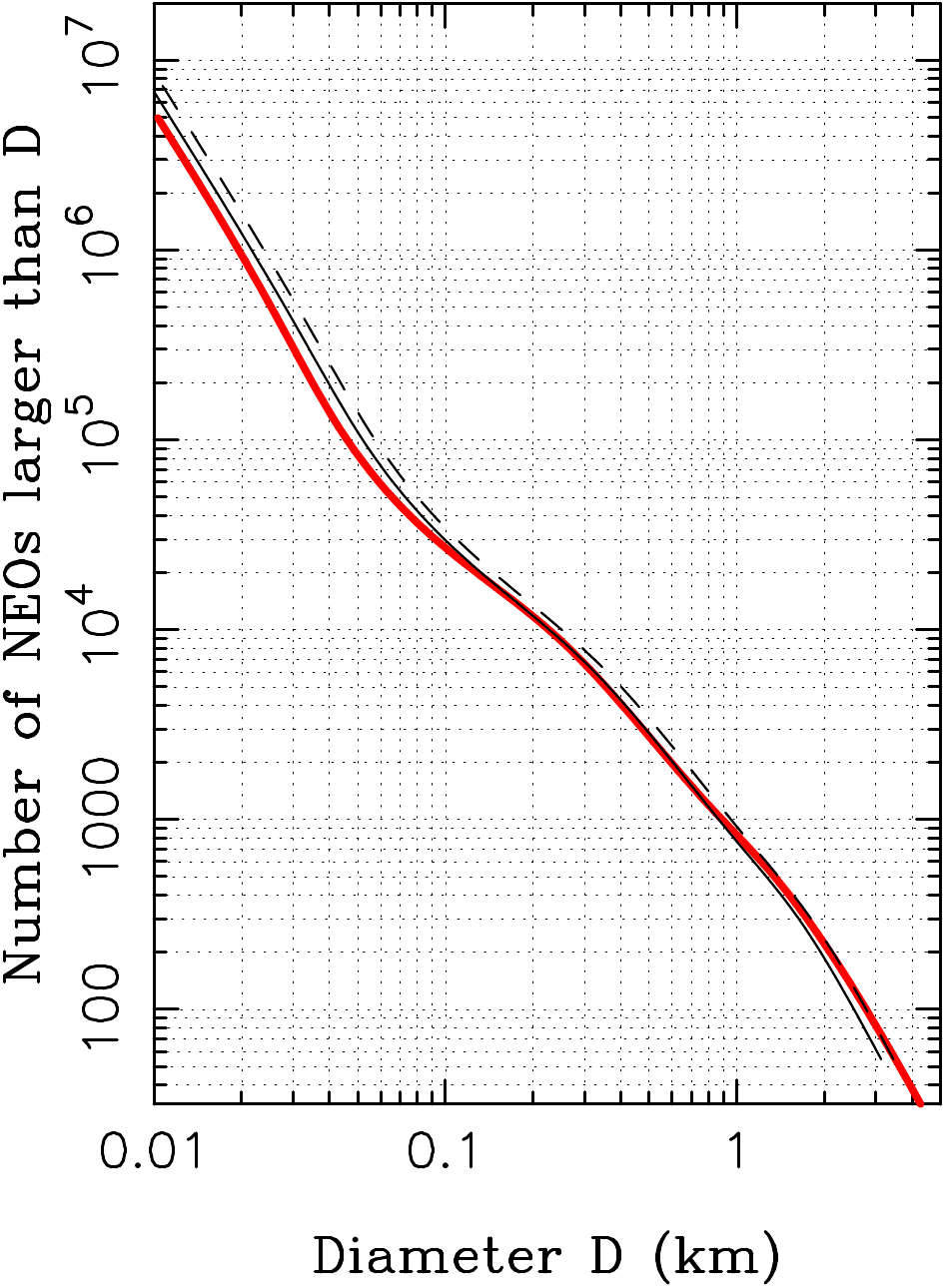}
\caption{The size distribution of NEOs from our size-dependent albedo model (red line) is compared to 
the size distributions constructed from the best-fit absolute magnitude distribution in Paper II and 
reference visual albedo $p_{\rm V}=0.14$ (dashed line) and 0.17 (solid line). The size-dependent albedo
distribution adopted here was constructed with the methods described in Sect. 4.1.}
\label{sizedist1}
\end{figure}

\begin{figure}
\epsscale{0.6}
\plotone{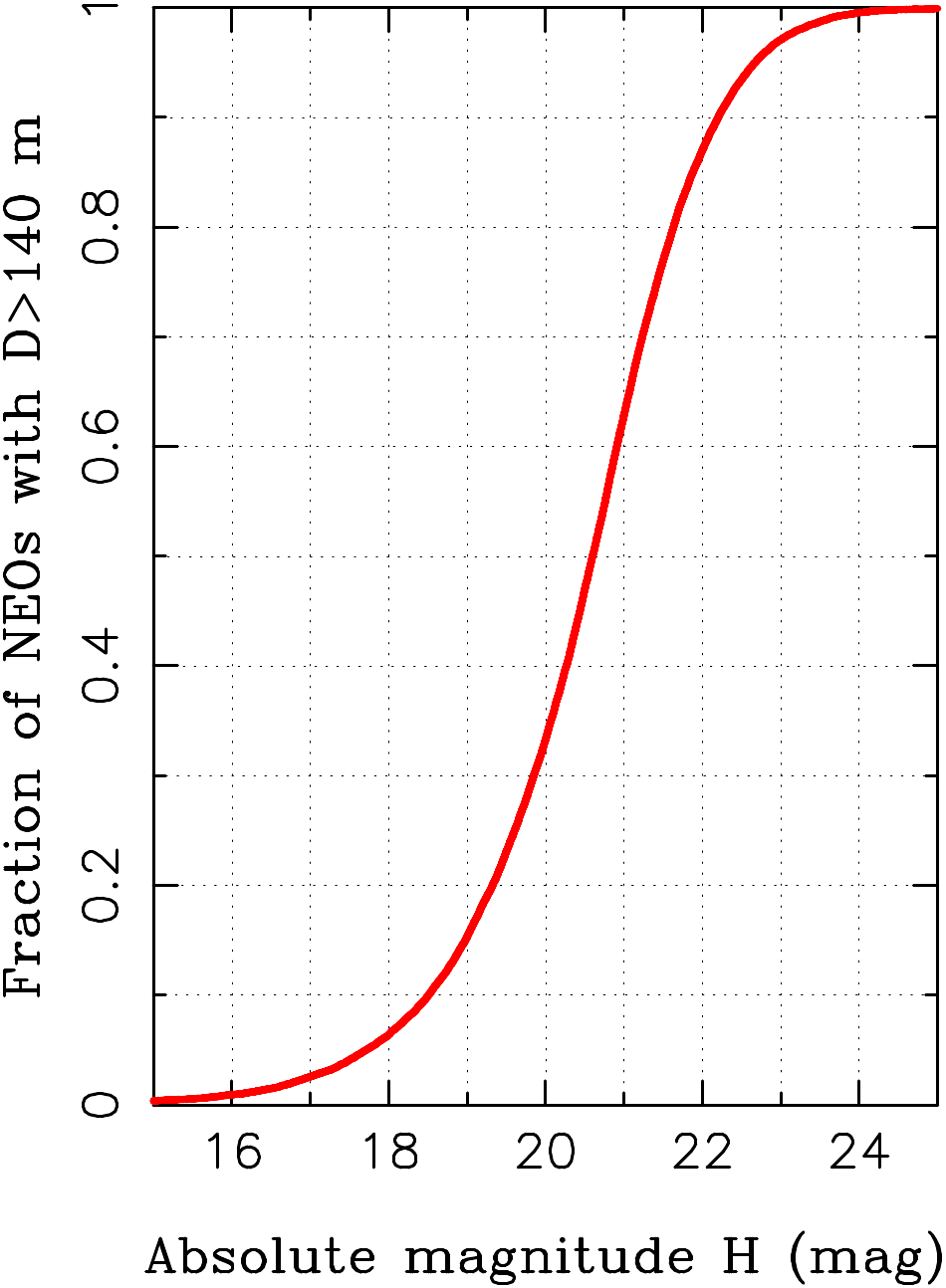}
\caption{The fraction of diameter $D>140$ m NEOs having magnitudes brighter than $H$. To reach 
a 90\% completion for $D>140$ m, telescopic observations in visible wavelengths would need to 
detect all NEOs brighter than $H=22.2$.}
\label{140}
\end{figure}

\begin{figure}
\epsscale{0.9}
\plotone{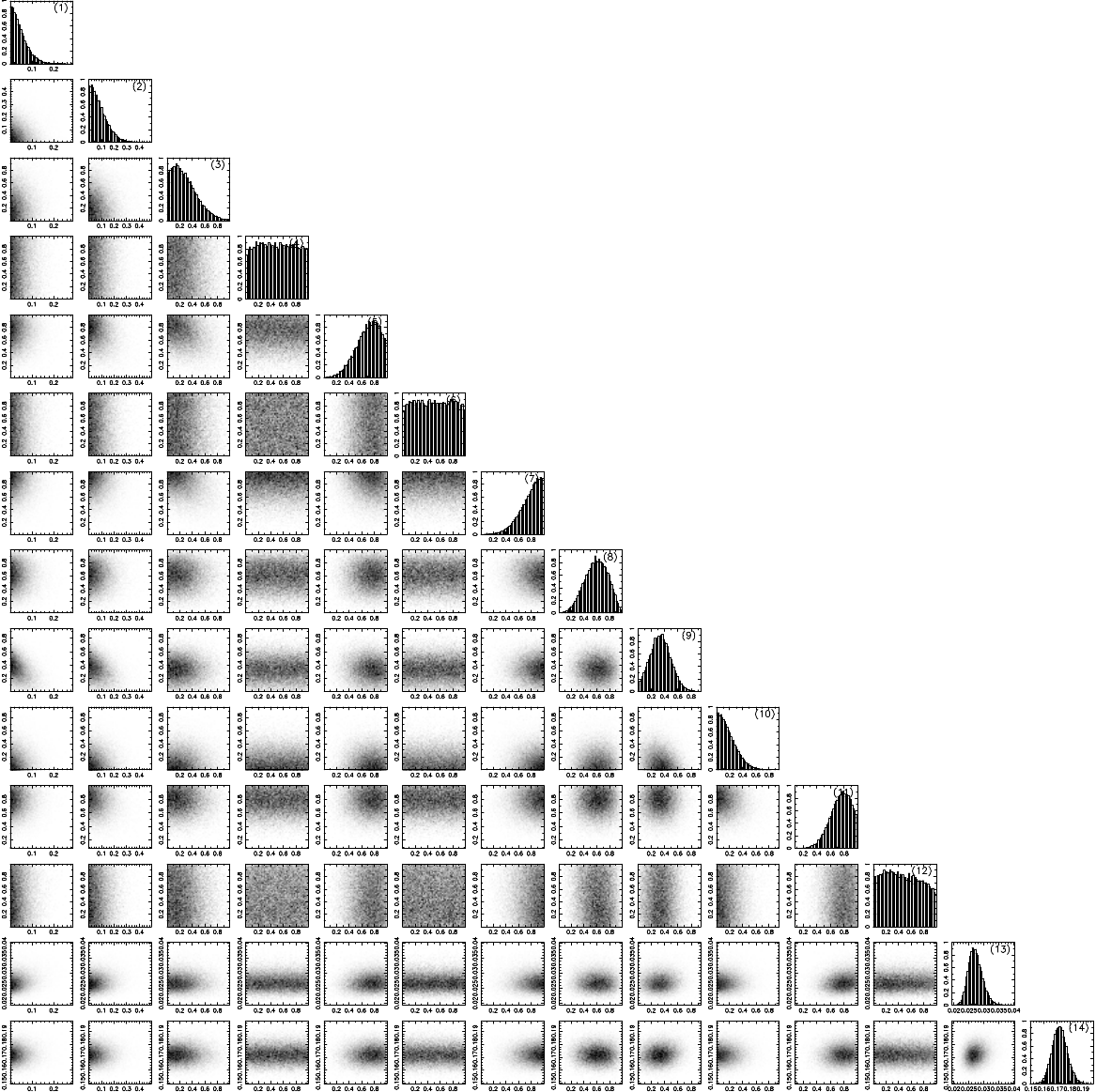}
\caption{The posterior distribution of 14 model parameters from our complex \texttt{MultiNest}
(Sects. 3.6). The individual plots are labeled (1) to (12) following the model parameter sequence 
given in Table 3.}
\label{triangle2}
\end{figure}

\begin{figure}
\epsscale{0.32}
\epsscale{1.0}
\plotone{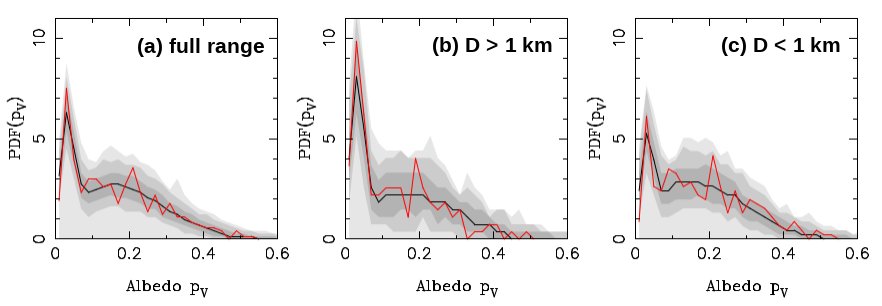}
\caption{The (biased) albedo distribution from the complex model is compared to NEOWISE detections.
From left to right the panels show the results for: (a) the full diameter range, (b) $D>1$ km, and 
(c) $D<1$ km. For $D>1$ km, the dark peak of the albedo distribution is prominent. For $D<1$ km, 
the dark peak is subdued, indicating that small NEOs more often have higher 
albedos. The model correctly reproduces the dependence of the albedo distribution on size.
}
\label{alb3}
\end{figure}

\begin{figure}
\epsscale{0.32}
\epsscale{1.0}
\plotone{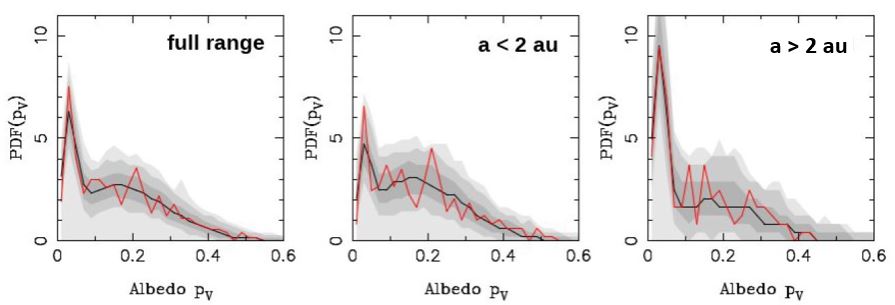}
\caption{The (biased) albedo distribution from the complex model is compared to NEOWISE detections.
From left to right the panels show the results for: (a) the full semimajor axis range, (b) $a<2$ au, and 
(c) $a>2$ au. For $a>2$ au, the dark peak of the albedo distribution is prominent. For $a<2$ au, 
the dark peak is subdued, indicating that NEOs with $a<2$ au more often have higher 
albedos. This happens because the inner belt sources have larger contributions to bright NEOs than
the outer belt sources. The model correctly reproduces the dependence of the albedo distribution
on the semimajor axis.}
\label{alb5}
\end{figure}

\begin{figure}
\epsscale{0.8}
\plotone{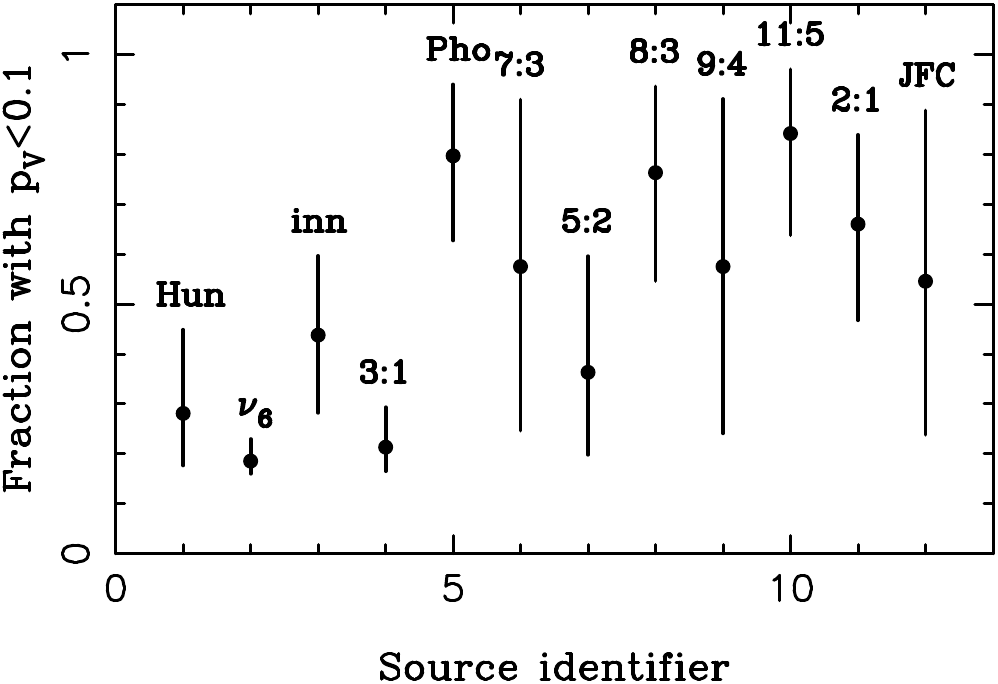}
\caption{The fraction of NEOs with $p_{\rm V}<0.1$ produced from each source. We used the $f_{{\rm d},j}$
parameters reported in Table 3 and summed up the contributions of dark and bright Rayleigh
distributions from each source to $p_{\rm V}<0.1$.} 
\label{type}
\end{figure}

\begin{figure}
\epsscale{0.32}
\epsscale{1.0}
\plotone{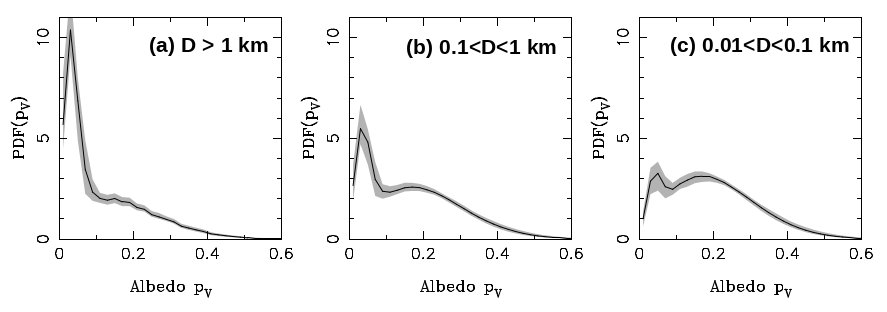}
\caption{The debiased albedo distribution of NEOs from the complex model.
From left to right the panels show the results for: (a) $D>1$ km, (b) $0.1<D<1$ km, and 
(c) $0.01<D<0.1$ km. For $D>1$ km, the dark peak of the albedo distribution is prominent. For $D<0.1$ km, 
the dark peak is subdued indicating that small NEOs are more often found in the peak with higher 
albedos. The mean albedos for the three size ranges shown here
are $\langle p_{\rm V} \rangle=0.124$ ($D>1$ km), 0.167 ($0.1<D<1$ km) and 0.191 ($0.01<D<0.1$ km).}
\label{dalb}
\end{figure}

\begin{figure}
\epsscale{0.6}
\plotone{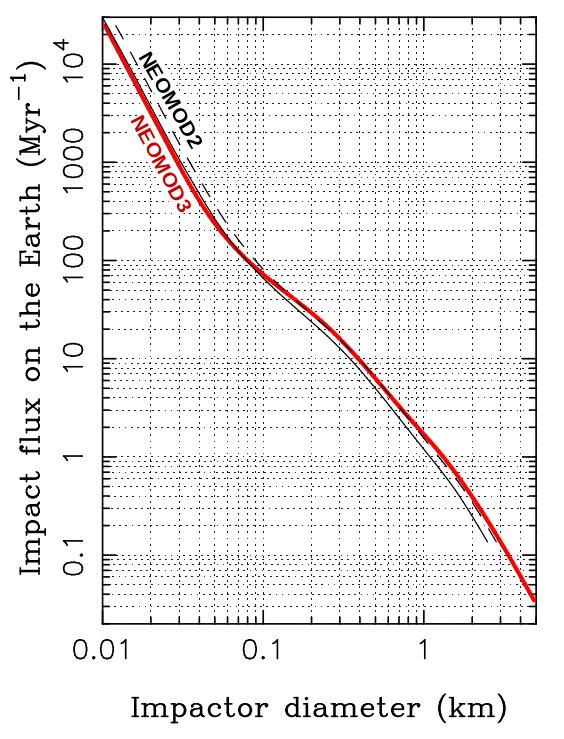}
\caption{The impact flux of NEOs on the Earth from our model with the size-dependent albedo distribution (red line). 
  The size-dependent albedo distribution model adopted here was constructed with the methods described in
  Sect. 4. The impact flux from NEOMOD2 is plotted for reference. For that we used $p_{\rm V}=0.14$ (dashed black line) 
  and $p_{\rm V}=0.18$ (solid black line) to translate the absolute magnitudes from NEOMOD2 to diameters.}
\label{impact}
\end{figure}

\begin{figure}
\epsscale{0.6}
\plotone{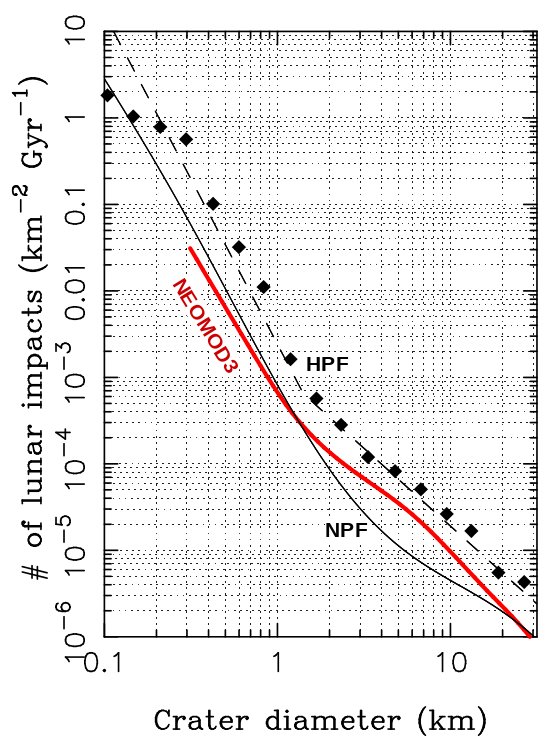}
\caption{The lunar production function. The plot shows the cumulative distribution of crater diameters 
produced on the lunar surface in Gyr. The number of craters is normalized to 1 km$^2$ of the lunar surface. 
The red line is the production function obtained here from NEOMOD (Sect. 5.5). The black solid line is the Neukum's 
production function (NPF) as reported in Table 1 in Ivanov et al. (2002) (the column ``New'' $N(D)$, 
Neukum et al. 2001). The black diamonds are the Hartmann's production function (HPF) from Hartmann (1995).
Finally, the dashed line is the piece-wise power-law fit to HPF as given in Eqs. (1a)-(1c) in 
Ivanov et al. (2002).}
\label{neukum}
\end{figure}

\begin{figure}
\epsscale{0.7}
\plotone{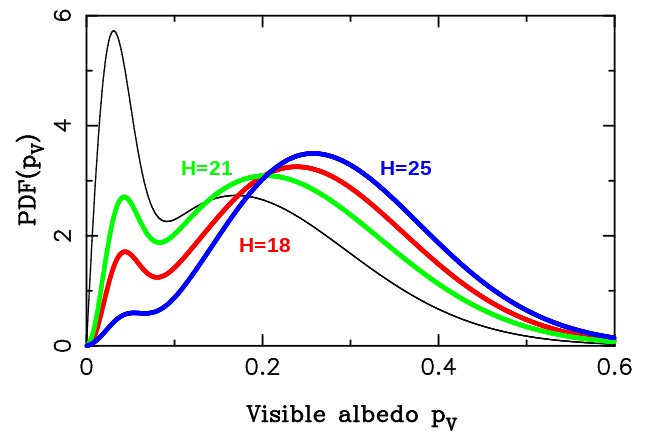}
\caption{The albedo distribution from our simple model, $\phi(p_{\rm V})$ (black line; Eq. (\ref{phi1}), Table~1)
is compared to the albedo distributions, $\phi'(p_{\rm V};H)$ (Eq. \ref{phi2}), for $H=18$ (red line), $H=21$ (green 
line) and $H=25$ (blue line). The difference between $\phi$ (size-based distribution) and $\phi'$ 
(absolute-magnitude-based distribution) is the largest for $H=25$, where the absolute magnitude distribution 
has the steepest slope 
(Fig. \ref{harris}).} 
\label{prime}
\end{figure}

In more general terms, $\phi(p_{\rm V})$ from Eq. (\ref{phi1}) and $\phi'(p_{\rm V})$ from Eq. (\ref{phi2})
are related via
\begin{equation}
\phi'(p_{\rm V};H) = C_{\rm N} {D \over \kappa} \psi(D) \phi(p_{\rm V};D)\ , 
\label{phis}
\end{equation} 
again with $\kappa = -5/\ln 10$, $D=D(H,p_{\rm V})= c 10^{-H/5}/\sqrt{p_{\rm V}}$, and $c=1329$ km. The right-hand 
side of Eq. (\ref{phis}) is to be evaluated for a fixed value of $H$. The normalization 
constant $C_{\rm N}$ assures that $\int {\rm d}p_{\rm V} \phi'(p_{\rm V};H)=1$ for any $H$ (also, by definition,  
$\int {\rm d}p_{\rm V} \phi(p_{\rm V};D)=1$ for any $D$). For a bimodal albedo distribution with $\phi(p_{\rm V})$ 
being represented by delta functions and a single power-law size distribution $\psi(D)$, Eq. (\ref{phis})
can be reduced to the arguments discussed above. Figure \ref{prime} illustrates a more general case where 
we adopt (size-independent) $\phi(p_{\rm V})$ from our simple model, the size distribution of NEOs shown in Fig. 
\ref{sizedist}, and compute $\phi'(p_{\rm V};H)$ from Eq. (\ref{phis}) for several different values 
of the absolute magnitude. The plot illustrates the difference between different definitions of albedo 
distribution.

\acknowledgements

\begin{center}
{\bf Acknowledgements} 
\end{center} 
\vspace*{-3.mm}
The simulations were performed on the NASA Pleiades Supercomputer. We thank the NASA NAS computing division 
for continued support. The work of DN, RD, and WFB was supported by the NASA Planetary Defense Coordination 
Office project ``Constructing a New Model of the Near-Earth Object Population''. DV acknowledges support
from the grant 23-04946S of the Czech Science Foundation. The work of SN, SRC, PWC and DF 
was conducted at the Jet Propulsion Laboratory, California Institute of Technology, under a contract with 
the National Aeronautics and Space Administration. 

\clearpage

\begin{table}
\centering
{
\begin{tabular}{lrrrr}
\hline \hline
label & parameter        & median & $-\sigma$ & $+\sigma$ \\                      
\hline                    
(1) & $f_{\rm d}$         & 0.233    & 0.028    & 0.030    \\ 
(2) & $d$                & 0.029   & 0.003    & 0.003     \\
(3) & $b$                & 0.170   & 0.006    & 0.006      \\ 
\hline \hline
\end{tabular}
}
\caption{The median and uncertainities of our simple (global) model parameters (Sect. 3.5).
The first column is the parameter/plot label in Fig. \ref{triangle}. The uncertainties 
reported here were obtained from the posterior distribution produced by \texttt{MultiNest}.}
\end{table}

\begin{table}
\centering
{
\begin{tabular}{lrrrr}
\hline \hline
parameter & $D$  range & value & $D$ range & value \\
          & (km)       &   & (km)      &      \\
\hline
& \multicolumn{2}{c}{\it constant albedo model} & \multicolumn{2}{c}{\it variable albedo model}\\
$N_{\rm ref}$  & --           & $777\pm24$ & --           & $813\pm24$ \\                
$\beta_1$    & 0.001--0.028 & $2.54\pm0.03$   & 0.001--0.026 &  $2.53\pm0.03$\\ 
$\beta_2$    & 0.028--0.044 & $2.73\pm0.03$   & 0.026--0.041 &  $2.75\pm0.03$ \\
$\beta_3$    & 0.044--0.278 & $1.50\pm0.02$   & 0.041--0.261 &  $1.50\pm0.02$ \\ 
$\beta_4$    & 0.278--0.876 & $1.85\pm0.03$   & 0.261--0.824 &  $1.72\pm0.02$ \\
$\beta_5$    & 0.876--1.389  & $1.86\pm0.06$  & 0.824--1.306  & $1.66\pm0.05$\\
$\beta_6$    & 1.389--30.00   & $2.63\pm0.09$ & 1.306--30.00   & $2.58\pm0.09$ \\
\hline \hline
\end{tabular}
}
\caption{The normalization parameter ($N_{\rm ref}$) and cumulative slopes ($\beta_j= {\rm d} \log N/{\rm d} \log D$) 
  from our (debiased) size distribution models of NEOs. The constant albedo model was obtained from the 
  global (simple) fit to all NEOWISE data. The variable (i.e., size-dependent) albedo model was constructed from
  fits in different size ranges (Sect. 4.1). The parameter $N_{\rm ref}$ calibrates the size distribution
  for $D>1$ km. Note that the actual size distribution is computed from cubic splines that smoothly connect
  the slopes in different segments (Figs. \ref{sizedist} and \ref{sizedist1}). The number of $D>1$ km NEOs
  is thus (slightly) larger than $N_{\rm ref}$ (Paper I).}
\end{table}

\begin{table}
\centering
{
\begin{tabular}{lrrr}
\hline \hline
                    & $N_1(D)$   & $N_2(D)$ & $N_3(D)$    \\ 
\hline                    
$D>1$ km            & 779        & 891      & 828      \\ 
$D>300$ m           & 7330       & 8208     & 6620     \\ 
$D>140$ m           & 20,000     & 22,100   & 18,000   \\
$D>100$ m           & 30,200     & 33,500   & 27,000   \\
$D>30$ m            & 368,000    & 427,000  & 307,000  \\
\hline \hline
\end{tabular}
}
\caption{The estimated number of NEOs, $N(D)$, larger than diameter $D$.
  In the second column, $N_1(D)$ stands for the estimates obtained from the global (simple) model
  to all NEOWISE data. In the third column, $N_2(D)$ is based on the fixed albedo model
  obtained for $1<D<3$ km and used for all NEOs. In the forth column, $N_3(D)$ corresponds
  to the variable albedo model constructed with the methods described in Sect. 4.1. The
  range of estimates given here roughly expresses the uncertainty related to the albedo
  distribution. The ranges given in the abstract and conclusions, $830\pm60$ NEOs  
  with $D>1$ km and $20,\!000 \pm 2,\!000$ NEOs with $D>140$ m, conservatively contain different 
  estimates from all models presented here, including the complex model results from Sect. 4.2.}
\end{table}

\clearpage

\begin{table}
\centering
{
\begin{tabular}{lrrrrr}
\hline \hline
label & parameter        & median & $-\sigma$ & $+\sigma$ & limit \\                      
\hline
(1) & $f_{\rm d}(\nu_6)$            & 0.037   & 0.025    & 0.044 & 0.054 \\ 
(2) & $f_{\rm d}($3:1)              & 0.069   & 0.048    & 0.080 & 0.103 \\
(3) & $f_{\rm d}($5:2)              & 0.247   & 0.165    & 0.232 & --    \\ 
(4) & $f_{\rm d}($7:3)              & 0.498   & 0.329    & 0.334 & --    \\ 
(5) & $f_{\rm d}($8:3)              & 0.721   & 0.217    & 0.173 & --     \\ 
(6) & $f_{\rm d}($9:4)              & 0.498   & 0.334    & 0.336 & --     \\                 
(7) & $f_{\rm d}($11:5)             & 0.814   & 0.204    & 0.129 & 0.723  \\   
(8) & $f_{\rm d}($2:1)              & 0.599   & 0.193    & 0.179 & --   \\   
(9) & $f_{\rm d}($inner)            & 0.335   & 0.156    & 0.159 & --   \\  
(10) & $f_{\rm d}($Hun)            & 0.149   & 0.104    & 0.169 & 0.222 \\ 
(11) & $f_{\rm d}($Pho)            & 0.761   & 0.170    & 0.143 & --   \\ 
(12) & $f_{\rm d}($comets)            & 0.463   & 0.308    & 0.342 & --  \\ 
(13) & $d$                         & 0.027   & 0.002    & 0.003 & --    \\
(14) & $b$                         & 0.172   & 0.006    & 0.006 & --     \\ 
\hline \hline
\end{tabular}
}
\caption{The median and uncertainities of our complex model parameters. The first column is the 
parameter/plot label in Fig. \ref{triangle2}. The uncertainties reported here were obtained from 
the posterior distribution produced by \texttt{MultiNest}. For parameters, for which the 
posterior distribution shown in Fig. \ref{triangle} peaks near zero, the last column reports 
the upper (68.3\% of posteriors fall between zero and that limit) or lower limit
(68.3\% of posteriors fall between that limit and one).}
\end{table}

\begin{table}
\centering
{
\begin{tabular}{lrrr}
\hline \hline
& $I_1(D)$   & $I_2(D)$ & $I_3(D)$    \\
& Myr$^{-1}$ & Myr$^{-1}$ & Myr$^{-1}$ \\
\hline                    
$D>1$ km            & 1.51     & 1.74     & 1.61      \\ 
$D>300$ m           & 16.1     & 18.1     & 14.4     \\ 
$D>140$ m           & 46.7     & 52.0     & 41.9   \\
$D>100$ m           & 72.7     & 81.0     & 64.8   \\
$D>30$ m            & 993      & 1160     & 829  \\
\hline \hline
\end{tabular}
}
\caption{The impact flux of NEOs on the Earth, $I(D)$, for bodies larger than diameter $D$.
  In the second column, $I_1(D)$ stands for the estimates obtained from the global (simple) model
  to all NEOWISE data. In the third column, $I_2(D)$ is based on the fixed albedo model
  obtained for $1<D<3$ km and used for all NEOs. In the forth column, $I_3(D)$ corresponds
  to the variable albedo model constructed with the methods described in Sect. 4.1. The
  range of estimates given here roughly expresses the uncertainty related to the albedo
  distribution.}
\end{table}

\clearpage

\clearpage

\clearpage



\clearpage

\clearpage

\clearpage

\clearpage

\clearpage

\clearpage

\clearpage

\clearpage

\clearpage


\clearpage

\clearpage

\clearpage

\clearpage


\clearpage


\begin{thebibliography}{}

\bibitem[Bierhaus et al.(2018)]{2018M&PS...53..638B} Bierhaus, E.~B. and 6 colleagues 2018.\ Secondary craters and ejecta across the solar system: Populations and effects on impact-crater-based chronologies.\ Meteoritics and Planetary Science 53, 638–671. doi:10.1111/maps.13057

\bibitem[Bottke et al.(2002)]{2002Icar..156..399B} Bottke, W.~F. and 6 colleagues 2002.\ Debiased Orbital and Absolute Magnitude Distribution of the Near-Earth Objects.\ Icarus 156, 399–433. doi:10.1006/icar.2001.6788

\bibitem[Brown et al.(2013)]{2013Natur.503..238B} Brown, P.~G. and 32 colleagues 2013.\ A 500-kiloton airburst over Chelyabinsk and an enhanced hazard from small impactors.\ Nature 503, 238–241. doi:10.1038/nature12741

\bibitem[Christensen et al.(2012)]{2012DPS....4421013C} Christensen, E. and 8 colleagues 2012.\ The Catalina Sky Survey: Current and Future Work.\ AAS/Division for Planetary Sciences Meeting Abstracts \#44.

\bibitem[DeMeo et al.(2009)]{2009Icar..202..160D} DeMeo, F.~E., Binzel, R.~P., Slivan, S.~M., Bus, S.~J.\ 2009.\ An extension of the Bus asteroid taxonomy into the near-infrared.\ Icarus 202, 160–180. doi:10.1016/j.icarus.2009.02.005
  
\bibitem[Feroz and Hobson(2008)]{2008MNRAS.384..449F} Feroz, F., Hobson, M.~P.\ 2008.\ Multimodal nested sampling: an efficient and robust alternative to Markov Chain Monte Carlo methods for astronomical data analyses.\ Monthly Notices of the Royal Astronomical Society 384, 449–463. doi:10.1111/j.1365-2966.2007.12353.x

\bibitem[Feroz et al.(2009)]{2009MNRAS.398.1601F} Feroz, F., Hobson, M.~P., Bridges, M.\ 2009.\ MULTINEST: an efficient and robust Bayesian inference tool for cosmology and particle physics.\ Monthly Notices of the Royal Astronomical Society 398, 1601–1614. doi:10.1111/j.1365-2966.2009.14548.x

\bibitem[Granvik et al.(2016)]{2016Natur.530..303G} Granvik, M. and 8 colleagues 2016.\ Super-catastrophic disruption of asteroids at small perihelion distances.\ Nature 530, 303–306. doi:10.1038/nature16934

\bibitem[Granvik et al.(2018)]{2018Icar..312..181G} Granvik, M. and 8 colleagues 2018.\ Debiased orbit and absolute-magnitude distributions for near-Earth objects.\ Icarus 312, 181–207. doi:10.1016/j.icarus.2018.04.018

\bibitem[Harris(1998)]{1998Icar..131..291H} Harris, A.~W.\ 1998.\ A Thermal Model for Near-Earth Asteroids.\ Icarus 131, 291–301. doi:10.1006/icar.1997.5865
  
\bibitem[Harris and D'Abramo(2015)]{2015Icar..257..302H} Harris, A.~W., D'Abramo, G.\ 2015.\ The population of near-Earth asteroids.\ Icarus 257, 302–312. doi:10.1016/j.icarus.2015.05.004

\bibitem[Harris and Chodas(2021)]{2021Icar..36514452H} Harris, A.~W., Chodas, P.~W.\ 2021 (HC21).\ The population of near-earth asteroids revisited and updated.\ Icarus 365. doi:10.1016/j.icarus.2021.114452

\bibitem[Harris and Chodas(2023)]{2021Icar..36514452I} Harris, A.~W., Chodas, P.~W.\ 2023.\ Update of NEA population and survey
  completion, ACM conference in Flagstaff, https://www.hou.usra.edu/meetings/acm2023/pdf/2519.pdf

\bibitem[Hartmann(1995)]{1995Metic..30..451H} Hartmann, W.\ 1995.\ Planetary cratering I: Lunar highlands and tests of hypotheses on crater populations.\ Meteoritics 30, 451. doi:10.1111/j.1945-5100.1995.tb01152.x

\bibitem[Ivanov et al.(2002)]{2002aste.book...89I} Ivanov, B.~A., Neukum, G., Bottke, W.~F., Hartmann, W.~K.\ 2002.\ The Comparison of Size-Frequency Distributions of Impact Craters and Asteroids and the Planetary Cratering Rate.\ Asteroids III 89–101.
  
\bibitem[JeongAhn and Malhotra(2014)]{2014Icar..229..236J} JeongAhn, Y., Malhotra, R.\ 2014.\ On the non-uniform distribution of the angular elements of near-Earth objects.\ Icarus 229, 236–246. doi:10.1016/j.icarus.2013.10.030
  
\bibitem[Liu et al.(2008)]{2008SPIE.7017E..0ML} Liu, F. and 12 colleagues 2008.\ Development of the Wide-field Infrared Survey Explorer (WISE) mission.\ Modeling, Systems Engineering, and Project Management for Astronomy III 7017. doi:10.1117/12.790087

\bibitem[Mainzer et al.(2005)]{2005SPIE.5899..262M} Mainzer, A.~K. and 7 colleagues 2005.\ Preliminary design of the Wide-Field Infrared Survey Explorer (WISE).\ UV/Optical/IR Space Telescopes: Innovative Technologies and Concepts II 5899, 262–273. doi:10.1117/12.611774
  
\bibitem[Mainzer et al.(2011)]{2011ApJ...743..156M} Mainzer, A. and 36 colleagues 2011.\ NEOWISE Observations of Near-Earth Objects: Preliminary Results.\ The Astrophysical Journal 743. doi:10.1088/0004-637X/743/2/156
  
\bibitem[Mainzer et al.(2019)]{2019PDSS..251.....M} Mainzer, A.~K. and 7 colleagues 2019.\ NEOWISE Diameters and Albedos V2.0.\ NASA Planetary Data System. doi:10.26033/18S3-2Z54

\bibitem[Marchi et al.(2009)]{2009AJ....137.4936M} Marchi, S., Mottola, S., Cremonese, G., Massironi, M., Martellato, E.\ 2009.\ A New Chronology for the Moon and Mercury.\ The Astronomical Journal 137, 4936–4948. doi:10.1088/0004-6256/137/6/4936

\bibitem[Marsset et al.(2022)]{2022AJ....163..165M} Marsset, M. and 12 colleagues 2022.\ The Debiased Compositional Distribution of MITHNEOS: Global Match between the Near-Earth and Main-belt Asteroid Populations, and Excess of D-type Near-Earth Objects.\ The Astronomical Journal 163. doi:10.3847/1538-3881/ac532f

\bibitem[Morbidelli et al.(2018)]{2018Icar..305..262M} Morbidelli, A. and 7 colleagues 2018.\ The timeline of the lunar bombardment: Revisited.\ Icarus 305, 262–276. doi:10.1016/j.icarus.2017.12.046

\bibitem[Morbidelli et al.(2020)]{2020Icar..34013631M} Morbidelli, A. and 7 colleagues 2020.\ Debiased albedo distribution for Near Earth Objects.\ Icarus 340. doi:10.1016/j.icarus.2020.113631

\bibitem[Muinonen et al.(1995)]{1995A&A...293..948M} Muinonen, K., Bowell, E., Lumme, K.\ 1995.\ Interrelating asteroid size, albedo, and magnitude distributions..\ Astronomy and Astrophysics 293, 948–952.
  
\bibitem[Naidu et al.(2017)]{2017DPS....4911204N} Naidu, S.~P., Chesley, S.~R., Farnocchia, D.\ 2017.\ Near-Earth Object Survey Simulation Software.\ AAS/Division for Planetary Sciences Meeting Abstracts \#49.

\bibitem[Nesvorn{\'y} et al.(2021)]{2021Icar..36814621N} Nesvorn{\'y}, D., Bottke, W.~F., Marchi, S.\ 2021.\ Dark primitive asteroids account for a large share of K/Pg-scale impacts on the Earth.\ Icarus 368. doi:10.1016/j.icarus.2021.114621

\bibitem[Nesvorn{\'y} et al.(2022)]{2022ApJ...941L...9N} Nesvorn{\'y}, D. and 6 colleagues 2022.\ Formation of Lunar Basins from Impacts of Leftover Planetesimals.\ The Astrophysical Journal 941. doi:10.3847/2041-8213/aca40e

\bibitem[Nesvorn{\'y} et al.(2023)]{2023AJ....166...55N} Nesvorn{\'y}, D. and 13 colleagues 2023 (Paper I).\ NEOMOD: A New Orbital Distribution Model for Near-Earth Objects.\ The Astronomical Journal 166. doi:10.3847/1538-3881/ace040

\bibitem[Nesvorn{\'y} et al.(2023)]{2023Icar..39915545N} Nesvorn{\'y}, D. and 6 colleagues 2023b.\ Early bombardment of the moon: Connecting the lunar crater record to the terrestrial planet formation.\ Icarus 399. doi:10.1016/j.icarus.2023.115545

\bibitem[Nesvorn{\'y} et al.(2024)]{2023AJ....166...55O} Nesvorn{\'y}, D. and 11 colleagues 2023 (Paper I). NEOMOD 2:
  An Updated Model of Near-Earth Objects from a Decade of Catalina Sky Survey Observations, Icarus, in press

\bibitem[Neukum et al.(2001)]{2001SSRv...96...55N} Neukum, G., Ivanov, B.~A., Hartmann, W.~K.\ 2001.\ Cratering Records in the Inner Solar System in Relation to the Lunar Reference System.\ Space Science Reviews 96, 55–86. doi:10.1023/A:1011989004263

\bibitem[Novakovi{\'c} et al.(2017)]{2017AJ....153..266N} Novakovi{\'c}, B., Tsirvoulis, G., Granvik, M., Todovi{\'c}, A.\ 2017.\ A Dark Asteroid Family in the Phocaea Region.\ The Astronomical Journal 153. doi:10.3847/1538-3881/aa6ea8
  
\bibitem[Pravec et al.(2012)]{2012Icar..221..365P} Pravec, P., Harris, A.~W., Ku{\v{s}}nir{\'a}k, P., Gal{\'a}d, A., Hornoch, K.\ 2012.\ Absolute magnitudes of asteroids and a revision of asteroid albedo estimates from WISE thermal observations.\ Icarus 221, 365–387. doi:10.1016/j.icarus.2012.07.026

\bibitem[Press et al.(1992)]{1992nrca.book.....P} Press, W.~H., Teukolsky, S.~A., Vetterling, W.~T., Flannery, B.~P.\ 1992.\ Numerical recipes in C. The art of scientific computing.\ Cambridge: University Press, 2nd edition

\bibitem[Russell(1916)]{1916ApJ....43..173R} Russell, H.~N.\ 1916.\ On the Albedo of the Planets and Their Satellites.\ The Astrophysical Journal 43, 173–196. doi:10.1086/142244

\bibitem[Stuart and Binzel(2004)]{2004Icar..170..295S} Stuart, J.~S., Binzel, R.~P.\ 2004.\ Bias-corrected population, size distribution, and impact hazard for the near-Earth objects.\ Icarus 170, 295–311. doi:10.1016/j.icarus.2004.03.018

\bibitem[Trilling et al.(2020)]{2020NatAs...4..940T} Trilling, D.~E. and 15 colleagues 2020.\ Spitzer's Solar System studies of asteroids, planets and the zodiacal cloud.\ Nature Astronomy 4, 940–946. doi:10.1038/s41550-020-01221-y

\bibitem[Wright et al.(2010)]{2010AJ....140.1868W} Wright, E.~L. and 37 colleagues 2010.\ The Wide-field Infrared Survey Explorer (WISE): Mission Description and Initial On-orbit Performance.\ The Astronomical Journal 140, 1868–1881. doi:10.1088/0004-6256/140/6/1868
 
\bibitem[Wright et al.(2016)]{2016AJ....152...79W} Wright, E.~L., Mainzer, A., Masiero, J., Grav, T., Bauer, J.\ 2016.\ The Albedo Distribution of Near Earth Asteroids.\ The Astronomical Journal 152. doi:10.3847/0004-6256/152/4/79

\end{thebibliography}
\end{document}